\documentclass[10pt,journal,compsoc]{IEEEtran}

\usepackage{booktabs} 

\usepackage{wrapfig}
\usepackage{multirow}
\usepackage{color}
\usepackage{graphicx}

\usepackage{subfig}

\usepackage{sidecap}
\usepackage{amssymb}
\usepackage{amsmath}
\usepackage{url}

\makeatletter
\let\NAT@parse\undefined
\makeatother
\usepackage{hyperref}  

\usepackage{fancyhdr}

\ifCLASSOPTIONcompsoc
  \usepackage[nocompress]{cite}
\else
  \usepackage{cite}
\fi

\ifCLASSINFOpdf
\else
\fi
\begin{document}

\title{A Two-part Transformer Network for Controllable Motion Synthesis}

\author{Shuaiying Hou,
        Hongyu Tao,
        Hujun Bao,
        and Weiwei Xu 
\IEEEcompsocitemizethanks{\IEEEcompsocthanksitem Shuaiying Hou, Hongyu Tao, Hujun Bao and Weiwei Xu are with the State Key Lab of CAD\&CG, Zhejiang University, China.\protect\\
E-mail: 11721044@zju.edu.cn, 3170102625@zju.edu.cn, bao@cad.zju.edu.cn and xww@cad.zju.edu.cn }}


\IEEEtitleabstractindextext{%
\begin{abstract}
\justifying
Although part-based motion synthesis networks have been investigated to reduce the complexity of modeling heterogeneous human motions, their computational cost remains prohibitive in interactive applications. To this end, we propose a novel two-part transformer network that aims to achieve high-quality, controllable motion synthesis results in real-time. Our network separates the skeleton into the upper and lower body parts, reducing the expensive cross-part fusion operations, and models the motions of each part separately through two streams of auto-regressive modules formed by multi-head attention layers. However, such a design might not sufficiently capture the correlations between the parts. We thus intentionally let the two parts share the features of the root joint and design a consistency loss to penalize the difference in the estimated root features and motions by these two auto-regressive modules, significantly improving the quality of synthesized motions. After training on our motion dataset, our network can synthesize a wide range of heterogeneous motions, like cartwheels and twists. Experimental and user study results demonstrate that our network is superior to state-of-the-art human motion synthesis networks in the quality of generated motions.
\end{abstract}
\par

\begin{IEEEkeywords}
Human motion synthesis, transformer, deep learning, heterogeneous motion, body parts \protect\\ \protect\\
© 2023 IEEE.  Personal use of this material is permitted.  Permission from IEEE must be obtained for all other uses, in any current or future media, including reprinting/republishing this material for advertising or promotional purposes, creating new collective works, for resale or redistribution to servers or lists, or reuse of any copyrighted component of this work in other works.
\end{IEEEkeywords}}


\maketitle

\IEEEdisplaynontitleabstractindextext

\IEEEpeerreviewmaketitle

\begin{figure*}[t]
  \centering
  \includegraphics[width=0.95\linewidth]{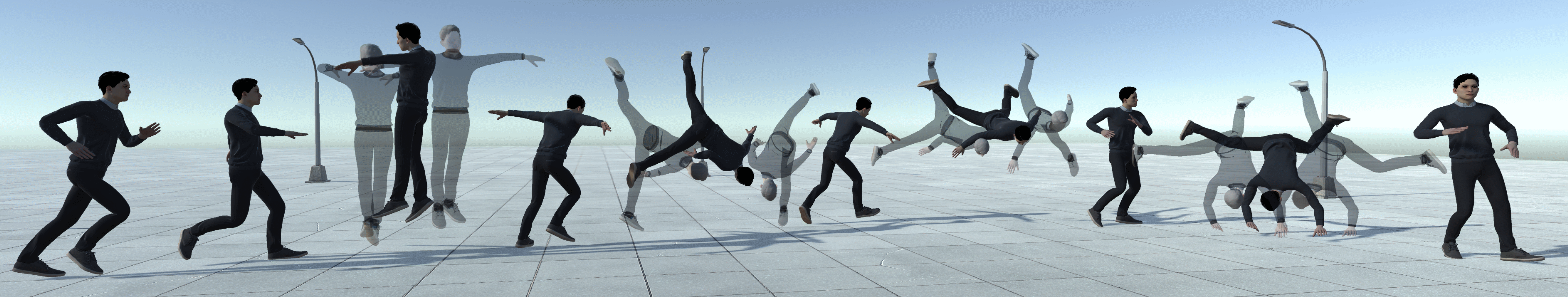}
  \caption{A motion sequence generated by our network. From left to right: running, twist, side somersault, whirlwind kicking, and ending with a cartwheel (from 0m27s to 0m37s in the accompanying video). Although the transitions between the last four types of heterogeneous motions are \textit{not} in the training dataset, our network can still successfully generate smooth transitions between them, which demonstrates the diversity of motions that our network can synthesize. The inverse kinematics algorithm is not applied in this experiment.}
  \label{fig:teaser}
\end{figure*}

\section{Introduction}
\label{sec:introduction}

Human motion synthesis has a wide range of applications in computer graphics and virtual reality, including video games, movie production, and crowd simulation. An appealing human motion synthesis method should scale up well to a large scope of motion types and generate realistic motions for virtual characters. Recently, deep neural networks (DNNs) have shown great potential to effectively model nonlinear kinematic processes embodied in different types of human motions, such as high-quality locomotion~\cite{holden2017phase, ling2020character}, contact-rich basketball motions~\cite{lee2018interactive, starke2020local}, martial arts movements~\cite{lee2021learning, lee2021learning2}, even motions with close multi-character interactions~\cite{starke2021neural}, and rhythmic dances~\cite{chen2020dynamic,perez2021transflower}.

Various neural network architectures have been developed to model the non-linearity in human motions~\cite{fragkiadaki2015recurrent, wang2019combining, holden2017phase, henter2020moglow}. The expert networks in the mixture-of-experts (MoE) architecture are used to cope with different subsets of input motions. Once trained, the experts become locally specialized for some subsets assigned by the gating networks, which helps MoE achieve satisfactory results on motion synthesis tasks~\cite{starke2019neural,ling2020character,starke2020local,starke2021neural}. Part-based motion synthesis networks adopt a divide-and-conquer strategy by breaking down the human body into parts based on its inherent skeleton structure~\cite{wang2019spatio, liu2021aggregated, ghosh2021synthesis}. Such approaches reduce the complexity of optimizing the entire human pose manifold, enabling these models to capture spatial variations better and avoid the over-smoothness of each part's motion, which can prevent generated motions from converging to mean poses. Moreover, these models improve the diversity of generated motions, such as frequent hand swings embodied in dances. However, as noted by Ghosh et al.~\cite{ghosh2021synthesis}, the quality of motions generated by part-based models still requires improvement in some aspects, such as foot slidings, limb constraints, and biomechanical plausibility. Another problem of existing part-based methods is that they are unsuitable for interactive control applications because of their high computational costs. Typically, the average depth and width of each part's network exceed the depth and width of the whole body's network divided by the number of parts, like the network presented in~\cite{liu2021aggregated}, resulting in increased computational complexity for part-based approaches compared to entire-body-based models. Furthermore, decomposing the entire body into more parts leads to extensive feature fusion operations. For instance, the methods presented in~\cite{wang2019spatio, ghosh2021synthesis} perform fusion and part-division operations seven times in the encoder and decoder, respectively. Unfortunately, modeling more parts and corresponding feature fusion operations reduces the network's parallelism, thereby further hindering its run-time performance.

In this paper, we propose a novel two-part transformer network (TPTN) to synthesize various types of high-quality motions conditioned on control signals in real-time. Such a two-part design is based on the observation that the upper body and lower body often move relatively independently depending on the types of motions. For instance, the upper body may perform different movements when the character walks or runs. In light of this, our approach divides the human body into two parts, the upper and lower body parts, and models them separately using two streams of auto-regressive modules (ARMs). This reduces the complexity of modeling the entire human pose manifold simultaneously and enables the synthesis of various types of heterogeneous motions. Compared to existing part-based motion synthesis methods, our two-part division significantly reduces the number of cross-part fusion operations, which benefits real-time performance. To address the potential loss of information caused by dividing the human body into two parts, we use a single linear layer for feature fusion to exchange information between the two ARMs. To improve correlations between the two body parts, we introduce consistency modules (CMs) to estimate the root motion from the shared root features and a consistency loss to penalize differences in the estimated root features and motions. The CMs are used to facilitate the consistency loss and are discarded after training to avoid incurring high computational costs.

Our TPTN takes control signals and motion features as inputs and outputs the conditioned probabilistic density function (PDF) of the next frame's motion. We also empirically keep relatively short frame buffers in the memory for the auto-regressive networks, which reduces computational costs in terms of time and space. The two-part and short frame-buffer design enables our TPTN to synthesize high-quality motions and respond to control signals in real-time. The architecture of our TPTN is based on the Transformer model~\cite{vaswani2017attention}, which has shown to be a powerful tool for modeling various types of data across different domains. We leverage multi-head attention layers in two ARMs to model the correlations and variations for the upper and lower body parts among a long-range temporal window. To make the TPTN have a finite temporal receptive field length and guarantee the temporal ordering of the motion sequence, we use the local self-attention mechanism  in~\cite{huang2020dance}. 

In summary, the contributions of our paper are as follows:
\begin{itemize}
    \item We propose a novel two-part auto-regressive neural network by adopting attention mechanisms to model the probability of human motion data. We demonstrate that it achieves state-of-the-art results and substantially improves the diversity of synthesized motions.
    \item We design a lightweight feature fusion layer and a consistency loss to help the network capture the correlations between the two parts' motions, which is fast to compute and crucial to the quality of generated motions.
\end{itemize}

The TPTN is a compact model of size \textasciitilde 3.59M bytes. Experimental results show that it can generate high-quality motions fast (\textasciitilde 75fps, \textasciitilde 63fps if inverse kinematic algorithm (IK) is applied) during inference. It can be easily applied to real-time applications such as video games or virtual live streaming. As shown in Fig.~\ref{fig:teaser}, our network can synthesize smooth motion transitions between complex heterogeneous motions. User studies also verify that the quality of motions generated by our network is superior to the motions of state-of-the-art human motion modeling and synthesis methods.

\begin{figure*}[t]
  \centering
    \subfloat[Auto-regressive modules and feature fusion layer]{
        \includegraphics[width=0.65\linewidth]{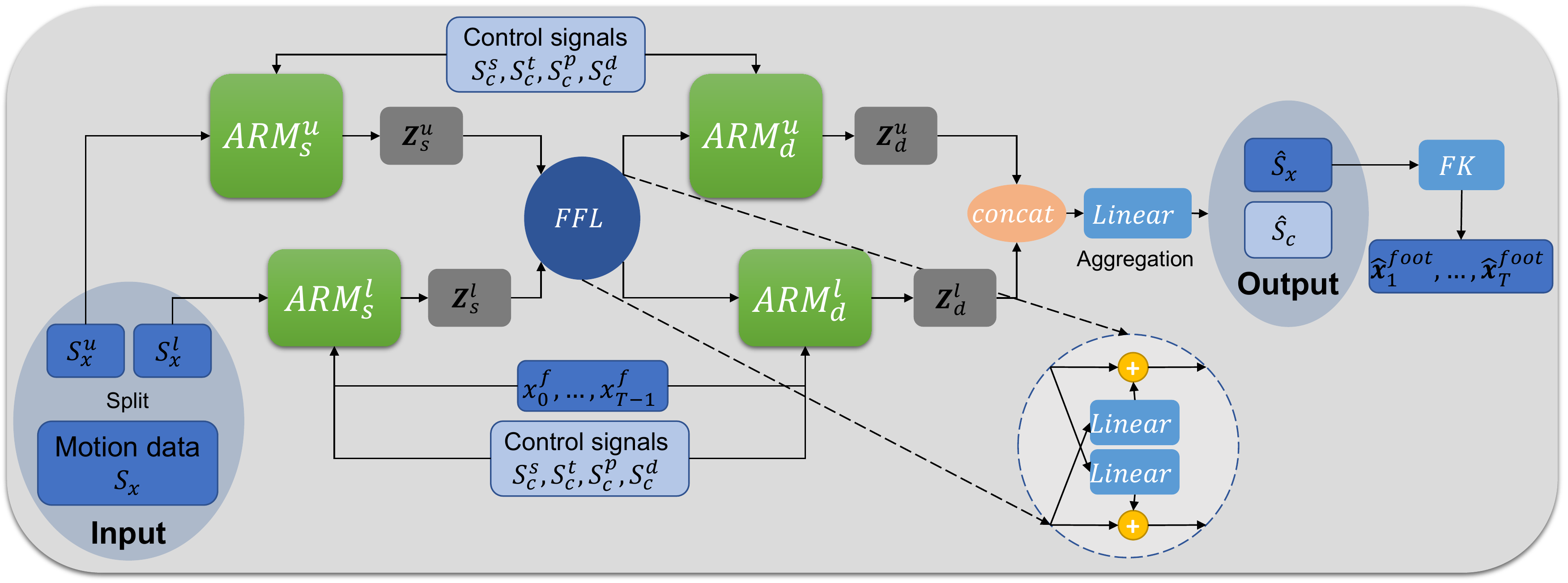}
        \label{fig:framework_main}
    }
    \subfloat[Foot-MLP and consistency modules]{
        \includegraphics[width=0.32\linewidth]{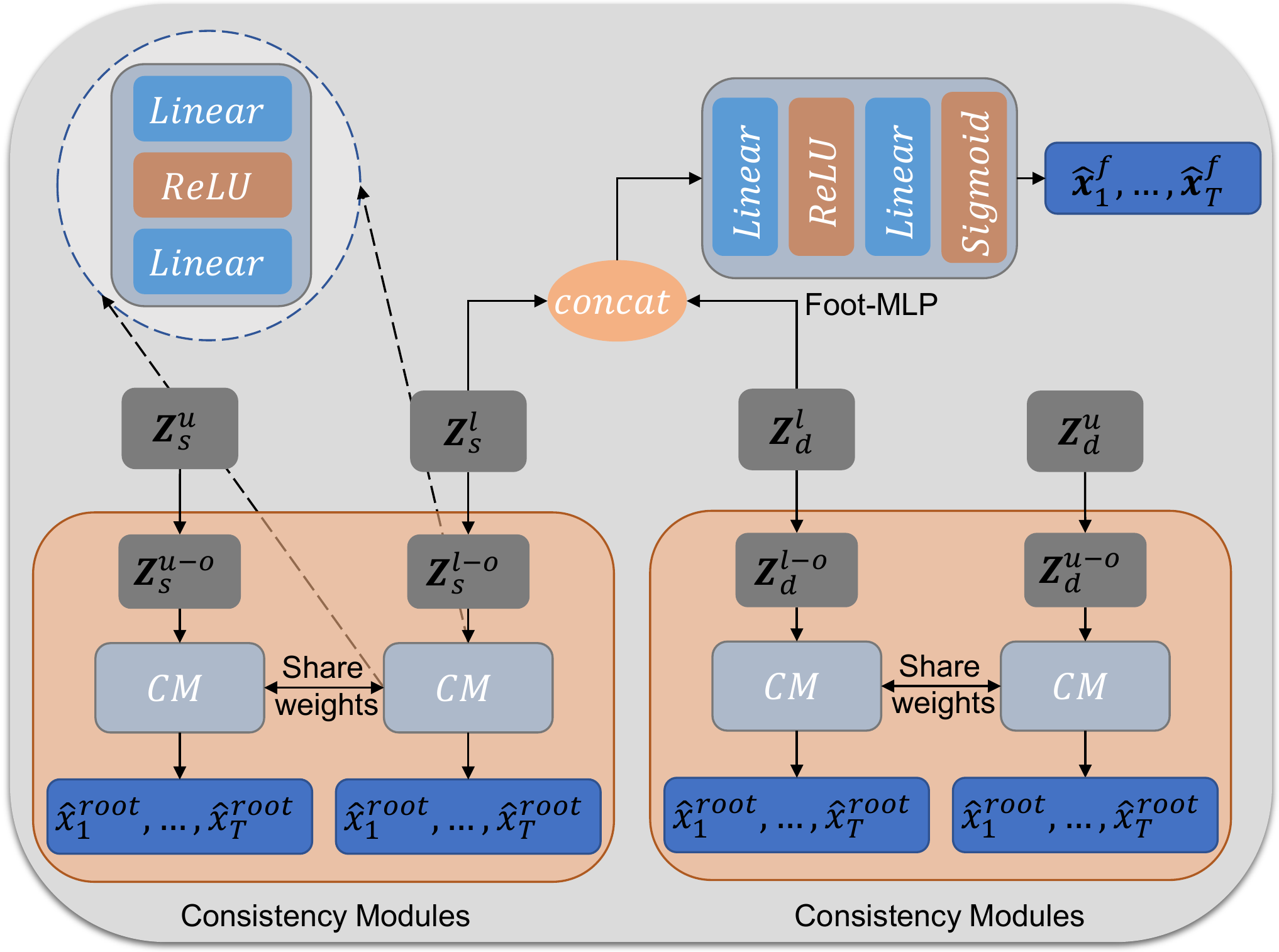}
        \label{fig:framework_cms}
    }
  \caption{The network architecture of the TPTN. (a) The upper and lower body parts are modeled by their own auto-regressive modules (ARMs). The linear aggregation layer then combines the two parts and produces the outputs for future frames, and the forward kinematics (FK) layer computes the foot positions of the output motion $\hat{S}_x$. (b) The Foot-MLP predicts the foot contact labels for future frames by taking the lower body features as input. The consistency modules (CMs) decode the overlapped features of the root back to the overlapped data $\hat{\mathbf{x}}_{n}^{root}$. Superscripts $u$ and $l$: ``upper body part'' and ``lower body part''. Subscripts $s$ and $d$: ``shallow layer'' and ``deep layer''. $\mathbf{Z}_{s}^u, \mathbf{Z}_{s}^l$ are feature sequences of the upper and lower body parts extracted by shallow layers, and $\mathbf{Z}_{d}^u, \mathbf{Z}_{d}^l$ by deep layers. $\mathbf{Z}_{s}^{u-o}, \mathbf{Z}_{s}^{l-o}$ are the overlapped feature sequences of $\mathbf{Z}_{s}^{u}, \mathbf{Z}_{s}^{l}$, and $\mathbf{Z}_{d}^{u-o}, \mathbf{Z}_{d}^{l-o}$ of $\mathbf{Z}_{d}^{u}, \mathbf{Z}_{d}^{l}$.}
  \label{fig:framework}
\end{figure*}

\section{Related Work}
\label{sec:related work}
With the development of motion capture techniques, statistical motion synthesis methods~\cite{xia2015realtime, starke2021neural, won2021control, aberman2020skeleton, liu2018learning} have become mainstream in recent decades. Please refer to~\cite{wang20143d} for a broad survey. We review deep learning-based motion synthesis methods mostly related to our work in the following.

An enormous number of deep learning-based methods have shown their capability of extracting powerful motion features to improve the motion synthesis results in recent years, which include recurrent neural networks~\cite{fragkiadaki2015recurrent, jain2016structural, Corona_2020_CVPR, martinez2017human, li2017auto, lee2018interactive}, fully connected networks~\cite{butepage2017deep, holden2016deep, holden2017phase}, graph networks~\cite{Li_2020_CVPR, Cui_2020_CVPR, yu2020structure, jang2022motion}, and generative adversarial networks (GAN)~\cite{villegas2018neural, liu2021aggregated, degardin2022generative, li2022ganimator}. Holden et al.~\cite{holden2016deep} learn a general motion manifold using a convolutional autoencoder and use another feed-forward network to disambiguate different control parameters. Aberman et al.~\cite{aberman2020skeleton} introduce a novel operation called skeletal convolution to process the skeleton's tree graph structure. Their system enables retargeting motions not observed in training. Ling et al.~\cite{ling2020character} and Starke et al.~\cite{starke2021neural} decouple motion synthesis and control signal synthesis into different modules to reduce the difficulty of training with large datasets. Ling et al.~\cite{ling2020character} model the motion manifold with a conditional variational auto-encoder (VAE) constructed by the mixture of experts (MoE) while generating control signals by different controllers in charge of different tasks using Reinforcement Learning. Starke et al.~\cite{starke2021neural} imitate animation layering in their control interface by layering the motion trajectories for different active behaviors generated by control modules, then generating motions from the edited motion trajectories employing an MoE motion generator. In contrast, we simultaneously predict future motions and control signals.

However, many deep learning-based methods suffer from over-smoothness or convergence towards mean poses when generating motions. Stochastic models show a way to avoid this problem by generating all possible motion sequences depending on the available information, like prior motion sequences, control signals, and random sampling noises. Fragkiadaki et al.~\cite{fragkiadaki2015recurrent} and Wang et al.~\cite{wang2019combining} propose to predict a conditional motion distribution represented by Gaussian Mixture Model (GMM). Sampling from the predicted GMM  can avoid the freezing issue in the generation of long motion sequences. Scheduled sampling~\cite{bengio2015scheduled} related techniques are also adopted~\cite{li2017auto,huang2020dance} for the same purpose, while applying such techniques will significantly slow down the training. Henter et al.~\cite{henter2020moglow} adopt normalizing flow in LSTM to get powerful and implicit conditional motion distributions, which greatly improves the diversity of synthesized motions. Introducing extra temporally-related variables called ``phase'' provides another option to cope with the over-smoothness problem. The phase variables can be classified into three categories: the contact-based global phase~\cite{holden2017phase} for cyclic motions like running or walking, the local phase~\cite{starke2020local} for acyclic motions like basketball motions, and the phase measured in the frequency domain that is automatically fitted in DeepPhase~\cite{starke2022deepphase} by neural networks. However, even with the DeepPhase method, different dimensions of phase manifolds still have to be fitted for different motions. In this paper, we adopt the attention strategy to model the motion as a Gaussian distribution similar to~\cite{fragkiadaki2015recurrent}.

The human skeleton is inherently tree-structured, and different body parts may influence human motion differently. Training in separate body parts and combining them back is similar to matching curves with piece-wise functions, making the fitting easier. Many methods ~\cite{wang2019spatio, liu2021aggregated, ghosh2021synthesis, jang2022motion, siyao2022bailando, tang2022real} have applied this idea to different motion synthesis tasks, such as motion prediction and style transfer. Wang et al.~\cite{wang2019spatio} decompose the human body into seven parts, which are encoded into a latent space through four fully-connected layers. Then they decode the motions of the seven body parts from the latent space and concatenate them to form motions for the entire human body. However, their method has a latency when controlling the motion synthesis because the prediction and control must be performed alternatively. Liu et al.~\cite{liu2021aggregated} construct five sub-GANs, each for one body part, and combine them through an aggregation layer forming the global GAN. Unfortunately, their AM-GAN has high computation cost in space (\textasciitilde 490M bytes parameters, \textasciitilde 5.5G GPU memory when inference), which hinders their application to online motion control. Ghosh et al.~\cite{ghosh2021synthesis} propose a hierarchical two-stream model to synthesize motions from textual descriptions. They split the human body into five parts, embed the hierarchical parts into an upper body and a lower body latent space separately, and then fed the latent spaces into two different GRUs. As they state in their paper, their model has a poor generalization ability, and the quality of generated motions can be improved further. These methods only perform the feature fusion between different body parts while do not constrain the correlations between them. In our work, the human body is divided into the upper and lower body parts to respond to control signals and better model heterogeneous motions effectively. The correlations between them are modeled by the feature fusion layer and consistency loss.

Recently, Transformer~\cite{vaswani2017attention} has bloomed in many tasks related to languages~\cite{brown2020language}, speeches~\cite{baevski2020wav2vec}, musics~\cite{dhariwal2020jukebox}, time-series sequences~\cite{zhou2021informer}, images~\cite{dosovitskiy2020image, liu2021swin}, and videos~\cite{yan2021videogpt} etc. Transformer has also been successfully applied to modeling human motions. Mao et al.~\cite{mao2020history} adopt attention mechanisms to compute the attention between the current motion context and the historical motion sub-sequences and achieve state-of-the-art motion prediction results on public human motion datasets. Researchers also used a cross-modal Transformer to synthesize dance motions conditioned on music~\cite{li2021ai, perez2021transflower, li2020learning, huang2020dance}. We also exploit the multi-head attention structure of the Transformer to capture the temporal correlations and variations contained in human motions.

\begin{figure}[t]
    \centering
    \subfloat[Skeletons of the performers]{
        \includegraphics[width=0.8\linewidth]{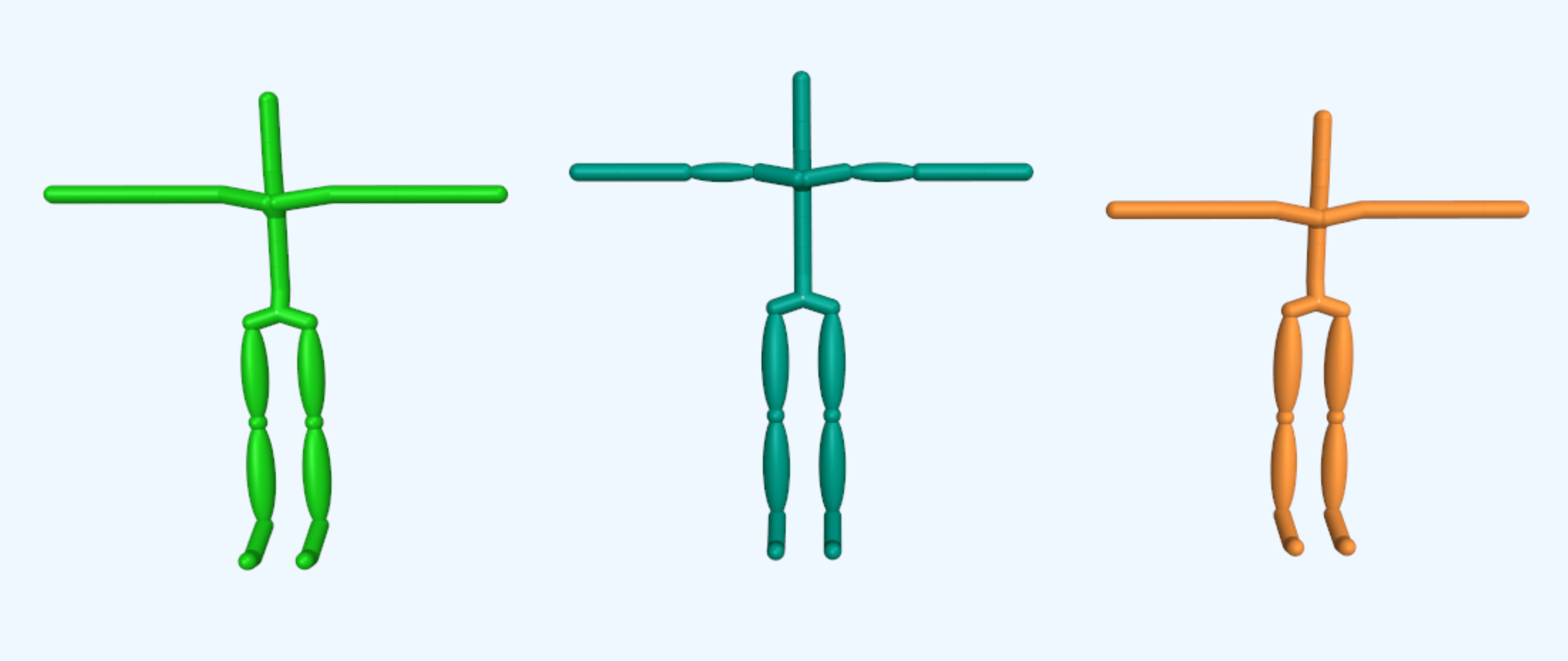}
        \label{fig:different sk}
    }
    \quad
    \subfloat[3D characters used for the performers]{
        \includegraphics[width=0.8\linewidth]{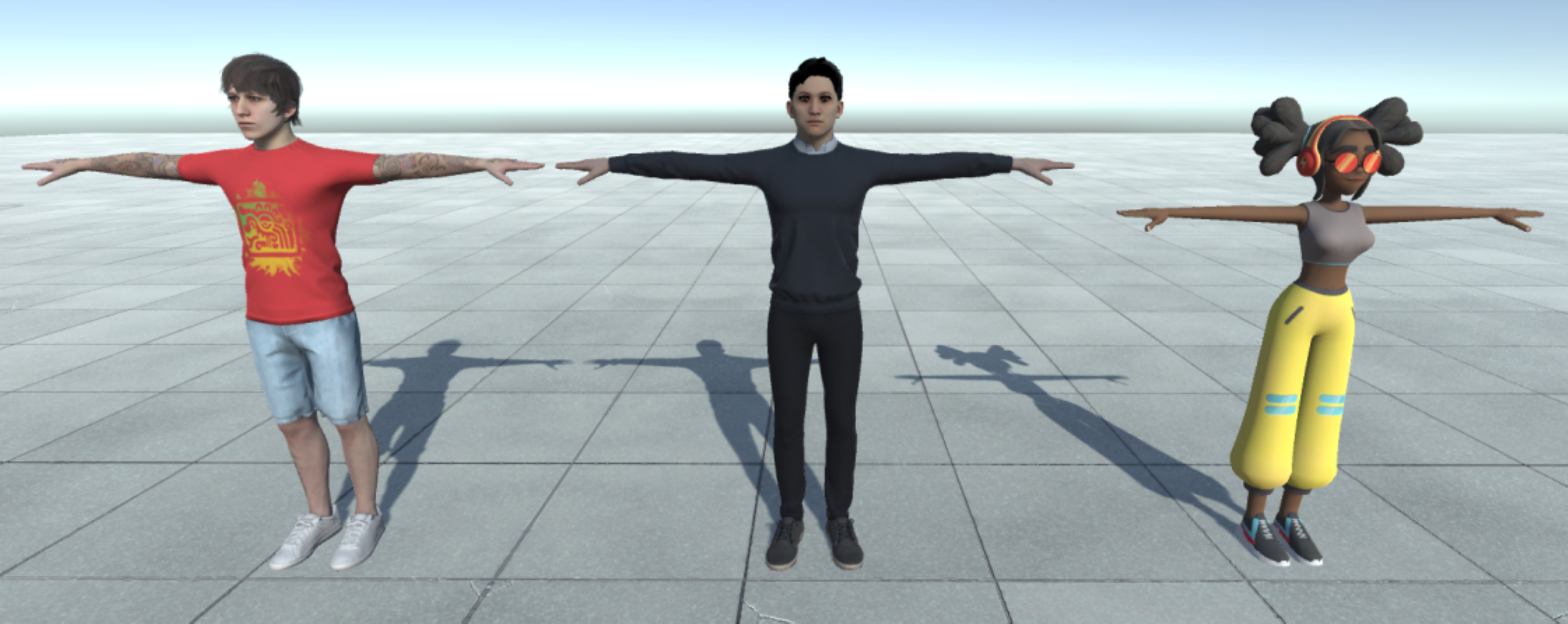}
        \label{fig:different mesh}
    }
    \caption{Skeletons and 3D characters in our dataset. The skeletons of the male dancer, male gymnast, and female dancer are from left to right. Their heights are 173cm, 181cm, and 166cm, respectively. The 3D characters are downloaded from Mixamo~\cite{adobe:mixamo}.}
    \label{fig:differentsk}
\end{figure}

\section{Data Processing and Representation}
\label{sec:dataset}
In this section, we first describe the settings of our motion dataset and then briefly describe the motion and control signal representations used in our network. Please refer to the supplementary material for the details of these two representations.

\noindent\textbf{Motion Dataset}.
Using the motion capture (mocap) technique, we build a complex heterogeneous human motion dataset of three performers. The performers consist of a male gymnast, a male dancer, and a female dancer, as shown in Fig.~\ref{fig:differentsk}. The dataset includes 21 types of motions: idle, walking, running, jumping with the left foot (jumping-L), jumping with the right foot (jumping-R), jumping with both feet (jumping-B), back walking (B-walking), kicking, punching, two kinds of whirlwind kicking (W-kicking and W-kicking-2), side somersault (S-somersault), twist, cartwheel, Chinese classical dance, modern dance, Xinjiang dance, Dai dance, Mongolian dance, Miao dance, and ballet.

The mocap data is recorded in BVH format in 120fps, and the skeleton for the character models is composed of $23$ joints in our system. Except for dancing motions, some transitions between different types of motions are also recorded. For some motion types, only one motion sequence is recorded. As a result, we can only train our model for these types of motions but cannot test the quality of generated motions. Therefore, we mirror these sequences and randomly put the original and mirrored ones into the training and test data separately. We down-sample the motion data to 60fps and obtain 172,956 and 37,390 frames in the training and test dataset.

\noindent\textbf{Motion Representation}.
\label{sec:data_process}
Our motion representation is similar to that in ~\cite{holden2017phase}. The motion representation vector $\mathcal{X}_{n} \in \mathbb{R}^{276}$ for the $n_{th}$ frame of motion consists of the root-joint information (the root angular velocities around the up axis, the root translational X and Z velocities, the root positions of Y-axis in global coordinate system, and the root rotations in the local root coordinate system), the non-root joint rotations, the non-root angular velocities, the 3D joint positions, and their linear velocities represented in the root's coordinate system. We use exponential maps~\cite{Grassia_1998} of the quaternion to represent the joint rotations. 

\noindent\textbf{Control Signal Representation}.
\label{sec:control_signal}
Our system supports four types of control signals, allowing the user to control how to synthesize motions in future frames. The control signals for the motion in the $n_{th}$ frame are denoted by $ \mathcal{C}_n = \{ \mathbf{c}_n^s, \mathbf{c}_n^t, \mathbf{c}_n^p, \mathbf{c}_n^d \}$, where $\mathbf{c}_n^s$ is the skeleton configuration represented by 3D joint positions at T-pose used to differentiate characters, $\mathbf{c}_n^t$ is the one-hot motion type to control the type of synthesized motions, $\mathbf{c}_n^p$ and $\mathbf{c}_n^d$ are the trajectory positions and directions in the future one second to control the forward path and direction of the synthesized motions. Similar to ~\cite{holden2017phase}, $\mathbf{c}_n^t$, $\mathbf{c}_n^p$ and $\mathbf{c}_n^d$ are uniformly sampled six frames within a one-second time window starting from the $n_{th}$ frame. The control signals used in training are extracted from the mocap data.

\section{Method}
We exploit the divide-and-conquer strategy by partitioning the human body into two parts - the upper and lower body parts- and structure our TPTN to learn features for different body parts to ease modeling heterogeneous motion data. In this section, we first describe the network structure of TPTN and then proceed to its training details.

\subsection{TPTN}
\label{sec:architechure}
Our TPTN $\mathcal{\psi}$ is a sequence-based auto-regressive model that contains four auto-regressive modules, $ARM_s^{u}$ for the upper body part, $ARM_s^{l}$ for the lower body part in shallow layers, and $ARM_d^{u}$, $ARM_d^{l}$ in deep layers. Each module is a transformer network with multi-head attention layers. An aggregation layer $\mathcal{\psi}_{Agg}$ is used to aggregate the two-part features before the final pose prediction as shown in Fig.~\ref{fig:framework_main} (detailed network parameters are reported in the supplementary material). The network is trained to model the Gaussian distribution of the predicted motion of the $n_{th}$ frame conditioned on the poses of previous frames and control signals with the following formula:

\begin{equation}
\begin{aligned}
&p(\mathcal{X}_{n}|\mathcal{X}_{i},...,\mathcal{X}_{n-1}, \mathcal{C}_{i},...,\mathcal{C}_{n-1}) = \mathcal{\psi}(\mathcal{X}_{i},...,\mathcal{X}_{n-1}, \mathcal{C}_{i},...,\mathcal{C}_{n-1})
\end{aligned}
\label{eq:motion_net}
\end{equation}

where $i=max(n-TRL, 0)$, and $TRL$ is the temporal receptive field length of the TPTN. Each frame in the output is conditioned on its previous $TRL$ frames at most. We let TPTN output the next frame for each input frame for the motion sequences in a batch to speed up the training. While during inference, TPTN predicts one future frame at each time step.

Given a sequence of $T$ frames motion data, $\mathbf{S}_x = \{ \mathcal{X}_{0},...,\mathcal{X}_{T-1}\}$, as input, the network $\mathcal{\psi}$ automatically splits each $\mathcal{X}_n$ in $\mathbf{S}_x$ into $\mathcal{X}_n^u$ for the upper body and $\mathcal{X}_n^l$ for the lower body, according to the joint indices. Note that we treat the foot contact labels $\mathbf{x}_n^f$ as part of $\mathcal{X}_n^l$. We intentionally let both $\mathcal{X}_n^u$ and $\mathcal{X}_n^l$ contain the root-joint information and call them the overlapped data $\mathbf{x}_{n}^{root}$. The TPTN models $\mathbf{S}_x^u = \{ \mathcal{X}_{0}^u,...,\mathcal{X}_{T-1}^u \}$ and $\mathbf{S}_x^l = \{ \mathcal{X}_{0}^l,...,\mathcal{X}_{T-1}^l \}$ separately through $ARM_s^{u}, ARM_d^{u}$ and $ARM_s^{l}, ARM_d^{l}$. The aggregation layer then combines the features output by $ARM_d^{u}$ and $ARM_d^{l}$ as the body-level feature to predict the motion Gaussian distribution $p(\mathcal{X}_{n}) = \mathcal{N}(\hat{\mu}_{n}, \hat{\sigma}_{n})$ in the next frame, where $\mathcal{X}_{n}$ is the motion representation vector, $\hat{\mu}_{n}$ its mean, and $\hat{\sigma}_{n}$ indicates the standard deviation computed by $\hat{\sigma}_{n} = e^{\sigma_{n}}$, where $\sigma_{n}$ is the direct output of the aggregation layer $\mathcal{\psi}_{Agg}$. This element-wise operation ensures that the standard deviation values in $\hat{\sigma}_{n}$ are always positive. Afterward, we can obtain the poses at frame $n$ by sampling from the predicted PDF (default setting in our experiments) or directly using the mean $\hat{\mu}_{n}$.

To ease the interactive control, we also train the proposed TPTN to output the next frame's trajectory positions and trajectory directions. Therefore, the final outputs of TPTN are $\hat{\mathbf{S}}_x = \{ (\hat{\mu}_{1}, \sigma_{1}),...,(\hat{\mu}_{T}, \sigma_{T}) \}$ and $\hat{\mathbf{S}}_c = \{ (\hat{c}_{1}^p, \hat{\sigma}_{1}^p, \hat{c}_{1}^d, \hat{\sigma}_{1}^d),...,(\hat{c}_{T}^p, \hat{\sigma}_{T}^p, \hat{c}_{T}^d, \hat{\sigma}_{T}^d) \}$, where $\hat{\sigma}_{n}^p$ and $\hat{\sigma}_{n}^d$ are the predicted standard deviations of the trajectory positions and directions computed the same way as $\hat{\sigma}_{n}$.

\begin{figure}[t]
  \centering
    \subfloat[Auto-regressive module for the upper body part]{
       \includegraphics[width=0.9\linewidth]{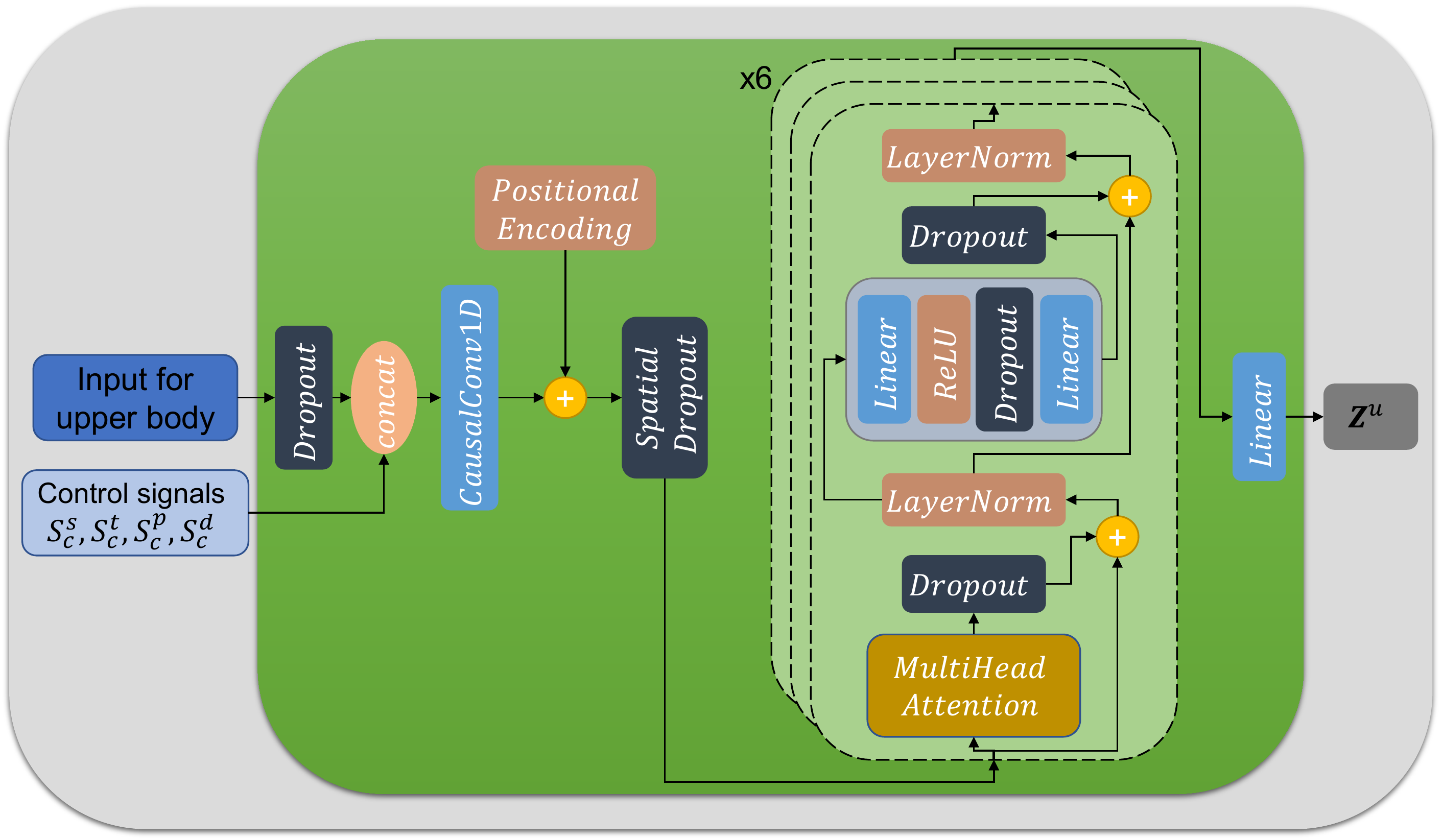}
       \label{fig:ar_u}
    }
    \quad
    \subfloat[Auto-regressive module for the lower body part]{
        \includegraphics[width=0.9\linewidth]{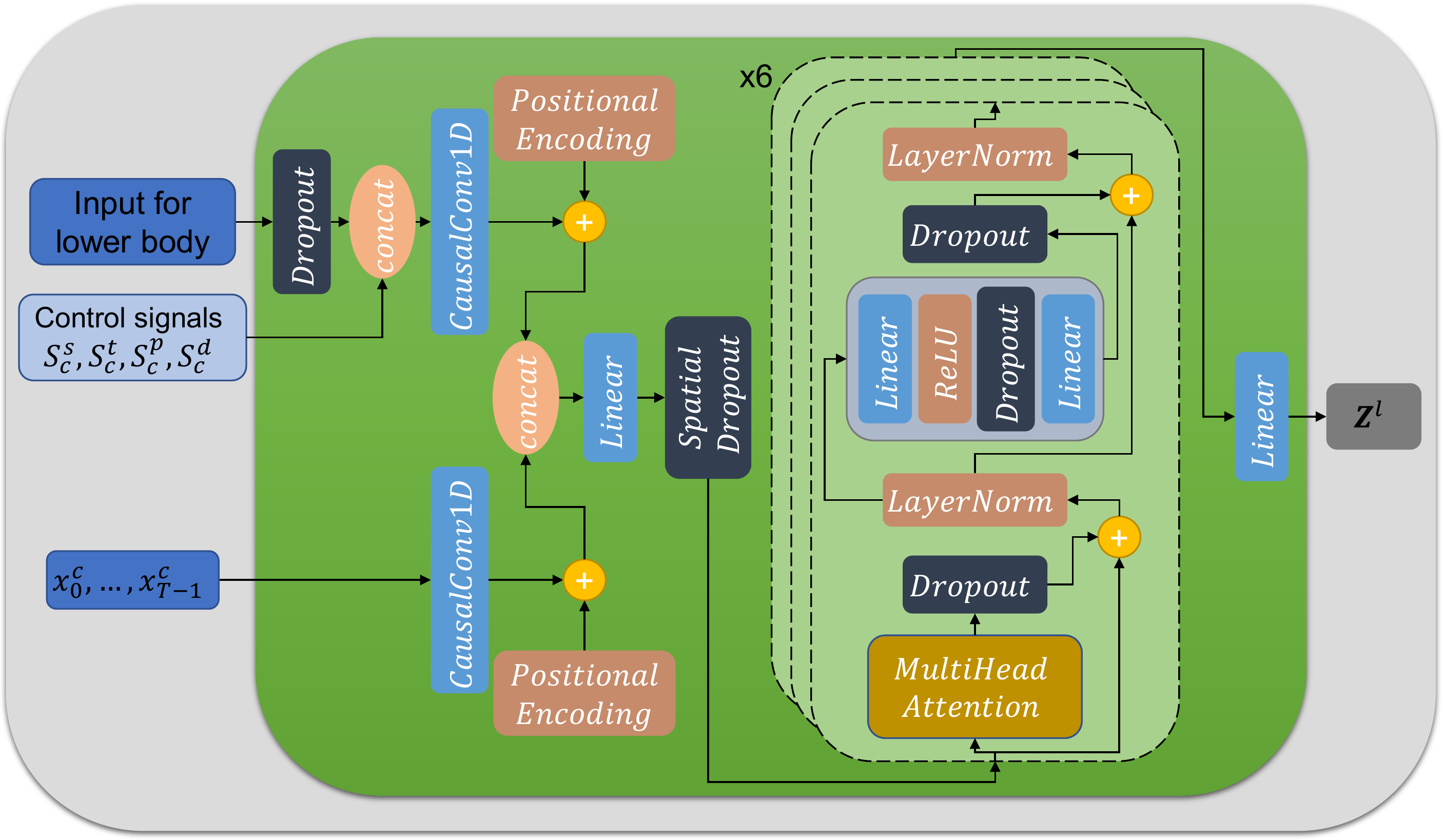}
        \label{fig:ar_l}
    }
  \caption{The detailed architecture of the auto-regressive modules for the upper and lower body parts in TPTN.}
  \label{fig:ars}
\end{figure}

\noindent\textbf{Auto-regressive Modules}. The auto-regressive modules for the upper and lower body parts are illustrated in Fig.~\ref{fig:ars}, and there are six multi-head attention layers in each ARM. The positional encoding and structure of multi-head attention layers are the same as that in~\cite {vaswani2017attention}. It is beneficial to convert the motion vectors into features before the multi-head attention layers to make the training more stable. Therefore, we use 1D causal convolutional layers, same as~\cite{oord2016wavenet}, to embed the input before adding the positional encodings. The 1D causal convolutions ensure the temporal ordering of the motion data, the kernel size $k_s$ of them for $ARM_s^{u}$ and $ARM_s^{l}$ is set to be three and $k_d$ for $ARM_d^{u}$ and $ARM_d^{l}$ be 1. We also add a dropout layer~\cite{JMLR:v15:srivastava14a} after the input and a 1D spatial dropout layer~\cite{tompson2015efficient} before the multi-head attention layers to resolve the possible over-fitting issue, and their drop probabilities are set to be 0.5 and 0.1, respectively. The control signals are concatenated to the input after dropout layers in each ARM to make the character responsive to user control signals. We set the local neighboring temporal window before the current frame $lm$ to be 8 in the local self-attention mechanism. As a result, $TRL$ of the TPTN is 87, which the following formula can compute:
\begin{equation}
TRL = (k_s-1) + (k_d-1) + lm + \sum_{i=2}^{6+6} (lm-1)
\end{equation}

\noindent\textbf{Feature Fusion Layer and Consistency Modules}. Different body parts are inherently correlated when the character performs motions, while independently modeling the body parts will weaken or even eliminate their correlations. To reserve the correlations, existing part-based models~\cite{wang2019spatio,liu2021aggregated,ghosh2021synthesis} fuse features of different body parts. Nevertheless, their methods are costly in time or space because of many cross-part fusion operations. Therefore, we fuse the features of the upper body part $\mathbf{Z}_{s}^u$ extracted by shallow layers $ARM_s^u$ and that of the lower body $\mathbf{Z}_{s}^l$ by $ARM_s^l$ to each other by the lightweight FFL shown in Fig.~\ref{fig:framework_main} only once in the middle of two ARMs. It makes our TPTN exchange information between the two body parts to extract more representative features and helps preserve the correlations better. Therefore, the features of high-level ARMs can be computed as follows:
\begin{subequations}
\begin{align}
\mathbf{Z}_{d}^u = ARM_d^{u}(\mathbf{Z}_{s}^u + Linear(\mathbf{Z}_{s}^l)) \\
\mathbf{Z}_{d}^l = ARM_d^{l}(\mathbf{Z}_{s}^l + Linear(\mathbf{Z}_{s}^u))
\end{align}
\end{subequations}

As described earlier in this section, we intentionally let $\mathbf{S}_x^u$ and $\mathbf{S}_x^l$ share the overlapped data. Hence, the latent features for the two parts computed by the ARMs contain encoded features of the root joint, termed overlapped features hereafter. Thanks to the self-attention mechanism, we can extract the overlapped features from the encoded features at the same positions as the transformer's input. To ensure that the overlapped features do represent the root joint, we then apply CMs to decode them back to the overlapped data $\hat{\mathbf{x}}_{n}^{root}$. A CM is a two-layer feed-forward network with ``ReLU'' activation. CMs are inserted at both the middle and final layers of the ARMs, and they will be discarded after training to reduce the computational cost of motion synthesis.

\noindent\textbf{Aggregation Layer and FK Layer}. The aggregation layer $Agg$ is a simple linear layer. It maps the concatenated features of $\mathbf{Z}_{d}^u$ and $\mathbf{Z}_{d}^l$ to the PDF of the predicted motion and outputs the trajectory position, and direction control signals for it. The FK layer performs the forward kinematics on the mean $\hat{\mu}_{n}$ of the output PDF and outputs the 3D foot positions $\hat{\mathbf{x}}_{n}^{foot}$ in the root's coordinate system. The FK layer is only activated during training.

\noindent\textbf{Foot-MLP}. The Foot-MLP is a two-layer feed-forward network with ``ReLU'' activation. It takes the lower body part's features $\mathbf{Z}_{s}^l$ and $\mathbf{Z}_{d}^l$ as input and predicts the binary foot contact labels for the predicted motion.

\subsection{Training Losses}
\label{sec:training_losses}

The training loss consists of six terms: the Gaussian loss $L_{G}$, the motion smoothness loss $L_{s}$, the foot contact loss $L_{f}$, the FK loss $L_{FK}$, the root consistency loss $L_{con}$, and the trajectory control loss $L_{t}$. It can be formulated into:
\begin{equation}
L = L_{G} + \lambda_{1}L_{s} + \lambda_{2}L_{f} + \lambda_{3}L_{FK} + \lambda_{4}L_{con} + \lambda_{5}L_{t}
\label{eq:training_loss}
\end{equation}
where the weight values of $\lambda_{1..5}$ are empirically set to $10.0$, $5.0$, $5.0$, $1.0$, and $1.0$ respectively in all our experiments.

\noindent\textbf{Gaussian loss}. This term is inspired by the Gaussian mixture loss in~\cite{fragkiadaki2015recurrent}, while we only use one mode and set the covariance matrix to be diagonal. It can be written into:
\begin{equation}
L_{G} = -\frac{1}{T} \sum_{n=0}^{T-1} ln(p(\mathcal{X}_{n}|\hat{\mu}_{n}, \hat{\sigma}_{n})),
\label{eq:gauss}
\end{equation}
where $\mathcal{X}_{n}$ is the motion representation vector extracted in the ${n}_{th}$ frame. The Gaussian loss term enforces the network to maximize the probability of the ground-truth mocap data during training. It is computed by inputting the ground-truth data into the predicted PDF. The covariance matrix is set to be diagonal in our implementation since it effectively reduces the number of parameters to speed up the training. For the correlations between different joint motions, we design an FK loss to implicitly capture them, which will be explained shortly.

We add a constraint to our implementation to ensure that the standard deviation $\hat{\sigma}_{n}$ is greater than 1e-4 by a clamping operation. After training, we observe that the standard deviation output by the trained TPTN is usually between 1e-3 and 1e-2. We hypothesize that the parameter $\hat{\sigma}_{n}$ can allow the network to adjust the accuracy of the predicted future pose adaptively.


\noindent\textbf{Smoothness loss}. This term is a soft constraint to prevent the sudden change of velocities at joints and make the synthesized motion smoother, which can be formulated as:
\begin{equation}
L_{s} = \frac{1}{T-2} \sum_{n=1}^{T-2} \parallel \hat{\mu}_{n-1} + \hat{\mu}_{n+1} - 2\hat{\mu}_{n} \parallel_2,
\end{equation}
The smoothness loss is only optimized for the mean of the predicted PDF since the motion generated by the network is usually close to the mean in each frame.

\noindent\textbf{Foot contact loss}. We adopt the binary cross-entropy (BCE) loss function to train the network to predict whether the foot is in contact with the supporting plane in the $n_{th}$ frame:
\begin{equation}
L_{f} = \frac{1}{T} \sum_{n=0}^{T-1} BCE(\mathbf{x}_n^f, \hat{\mathbf{x}}_n^f),
\end{equation}
where $\hat{\mathbf{x}}_n^f$ are the predicted foot contact labels, which can be used to trigger IK algorithms to remove the foot slidings in the synthesized motion.

\noindent\textbf{FK loss}. We add an FK layer to the network to use the mean of predicted PDF to compute the 3D foot positions $\hat{\mathbf{x}}_{n}^{f}$ in the root's coordinate system, and then compute the FK loss to alleviate the foot sliding in generated motions:
\begin{equation}
\begin{aligned}
  L_{FK} = & \frac{1}{T} \sum_{n=0}^{T-1} { \parallel \hat{\mathbf{x}}_{n}^{foot} - {\mathbf{x}}_{n}^{foot} \parallel_2}  + \\
  & \frac{1}{T-1} \sum_{n=1}^{T-1} {\hat{\mathbf{x}}_{n-1}^f * \hat{\mathbf{x}}_{n}^f \parallel \hat{\mathbf{x}}_{n-1}^{foot} - \hat{\mu}_{n}^{foot} \parallel_1},
  \end{aligned}
  \label{eq:fkloss}
\end{equation}
where $\hat{\mathbf{x}}_{n-1}^f$ and $\hat{\mathbf{x}}_{n}^f$ are the predicted foot contact labels by Foot-MLP, $\hat{\mathbf{x}}_{n-1}^{foot}$ and $\hat{\mathbf{x}}_{n}^{foot}$ are the predicted foot positions by FK layer, and ${\mathbf{x}}_{n}^{foot}$ are the ground-truth foot positions of the $n_{th}$ frame. Since this term computes the foot positions with related joint angles, it can implicitly encode the correlation between joint angles on the leg.


The first term in $L_{FK}$ penalizes the differences between the predicted and the ground-truth foot positions. The second term makes the feet stick in the same position when the predicted foot contact labels of two adjacent frames are all ones to prevent foot sliding. In practice, the first term plays a major role in the beginning since its value is around 3\textasciitilde10 times larger than the second term. As the training proceeds, the second term contributes more to the optimization until it exceeds the first term at around the $50_{th}$ epoch, while foot contact loss $L_f$ has decreased from around 2.4 to around 0.3. Consequently, we observe that the second term tends to penalize the foot sliding to cooperate with $L_f$, and the optimization still succeeds in our experiments without disconnecting $L_{f}$.

\noindent\textbf{Root consistency loss}. We design this term to improve the learning of correlations between two body parts. It is formulated as follows:
\begin{equation}
L_{con} = L_{feat} + 2L_{root}
\label{eq:consistency_loss}
\end{equation}
where the first term $L_{feat}$ penalizes the difference between the overlapped features of the two body parts in the same feature level as follows:
\begin{equation}
  \begin{aligned}
  L_{feat} = \frac{1}{T} \sum_{n=0}^{T-1} \big( 
   & \parallel \mathbf{Z}_{s}^{u-o} - \mathbf{Z}_{s}^{l-o} \parallel_2 + \\
   & \parallel \mathbf{Z}_{d}^{u-o} - \mathbf{Z}_{d}^{l-o} \parallel_2
   \big)
  \end{aligned}
  \label{eq:feature_consistency}
\end{equation}
and the second term $L_{root}$ enforces that the overlapped features can be decoded back to the overlapped data by CMs. It penalizes the difference between the decoded and ground-truth overlapped data using the following formula:
\begin{equation}
  \begin{aligned}
  L_{root} = \frac{1}{T} \big(
   & \sum_{n=0}^{T-1} \parallel CM(\mathbf{Z}_{s}^{u-o}) - \mathbf{S}_x^{root} \parallel_2 + \\
   & \sum_{n=0}^{T-1} \parallel CM(\mathbf{Z}_{s}^{l-o}) - \mathbf{S}_x^{root} \parallel_2 + \\
   & \sum_{n=0}^{T-1} \parallel CM(\mathbf{Z}_{d}^{u-o}) - \mathbf{S}_x^{root} \parallel_2 + \\
   & \sum_{n=0}^{T-1} \parallel CM(\mathbf{Z}_{d}^{l-o}) - \mathbf{S}_x^{root} \parallel_2
   \big)
  \end{aligned}
\label{eq:root_consistency}
\end{equation}
where $\mathbf{Z}_{s}^{u-o}$ and $\mathbf{Z}_{s}^{l-o}$ are the overlapped features of $\mathbf{Z}_{s}^{u}$ and $\mathbf{Z}_{s}^{l}$, $\mathbf{Z}_{d}^{u-o}$ and $\mathbf{Z}_{d}^{l-o}$ are the overlapped features of $\mathbf{Z}_{d}^{u}$ and $\mathbf{Z}_{d}^{l}$, and $\mathbf{S}_x^{root} = \{ \mathbf{x}_{1}^{root},..., \mathbf{x}_{n}^{root} \}$ are the ground-truth overlapped data.

\begin{figure*}[t]
  \centering
    \subfloat[Cartwheels, S-somersault and Wkicking-
2]{
       \includegraphics[width=0.357\linewidth]{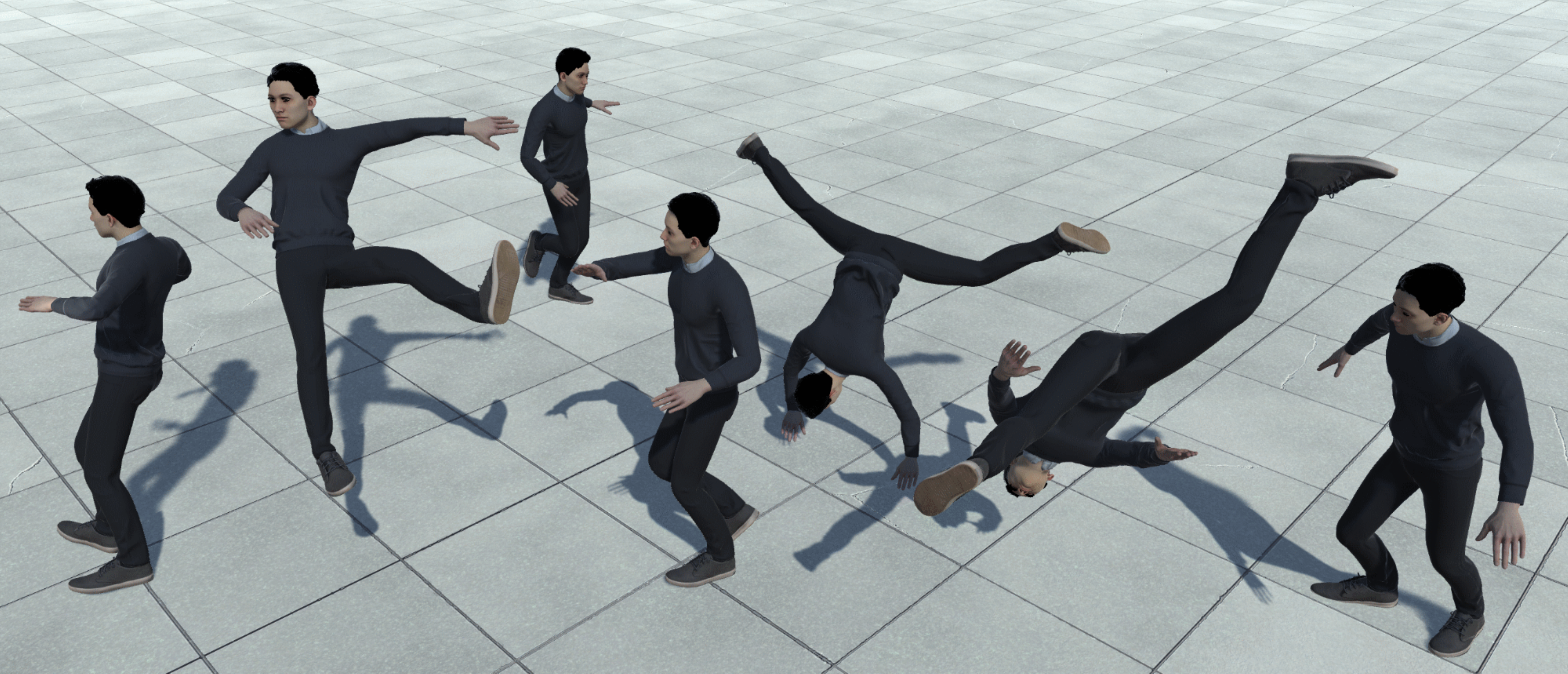}
       \label{fig:flips}
    }
    \quad
    \subfloat[Ballet]{
       \includegraphics[width=0.246\linewidth]{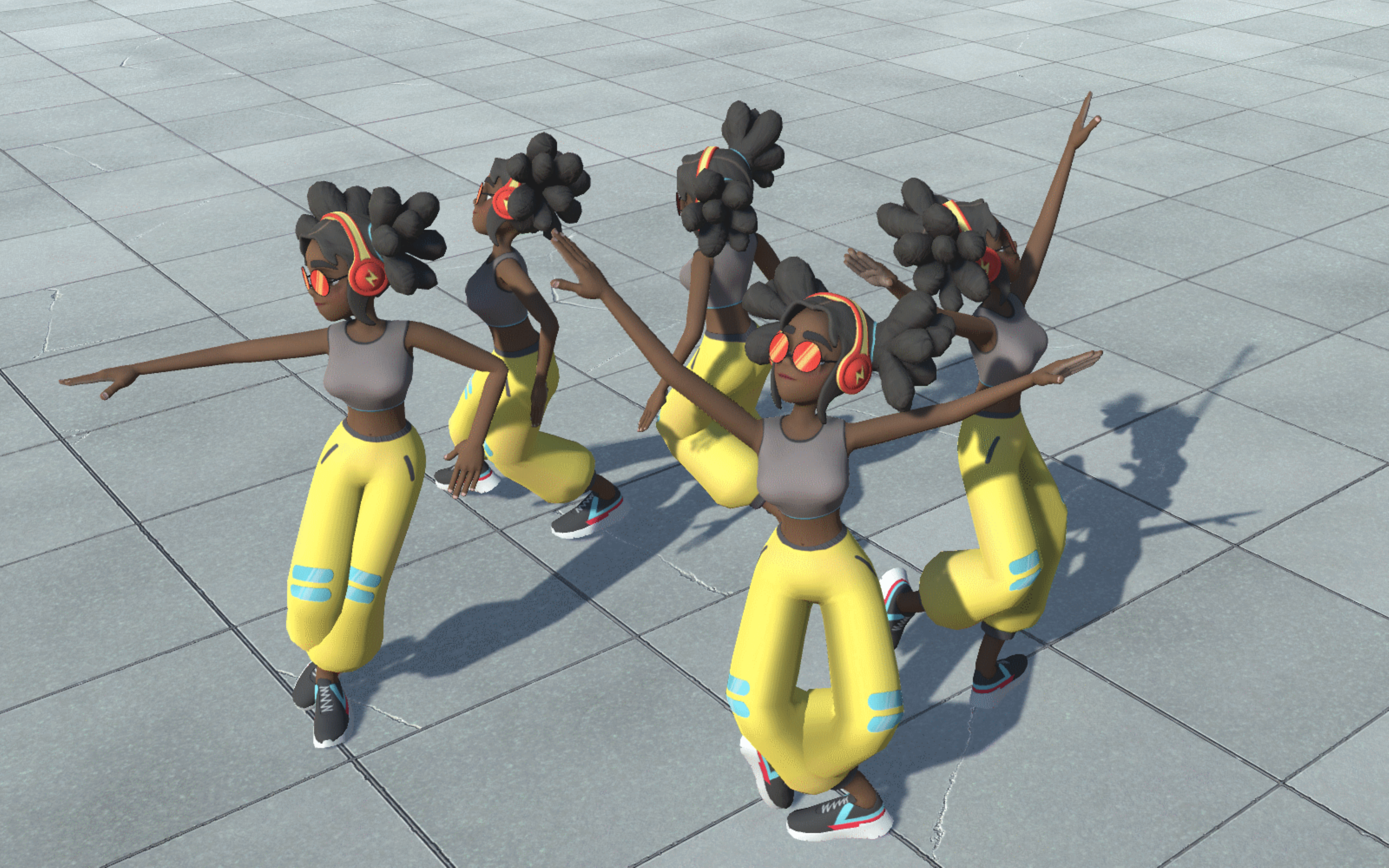}
       \label{fig:dance_female}
    }
    \quad
    \subfloat[Dai dance]{
        \includegraphics[width=0.305\linewidth]{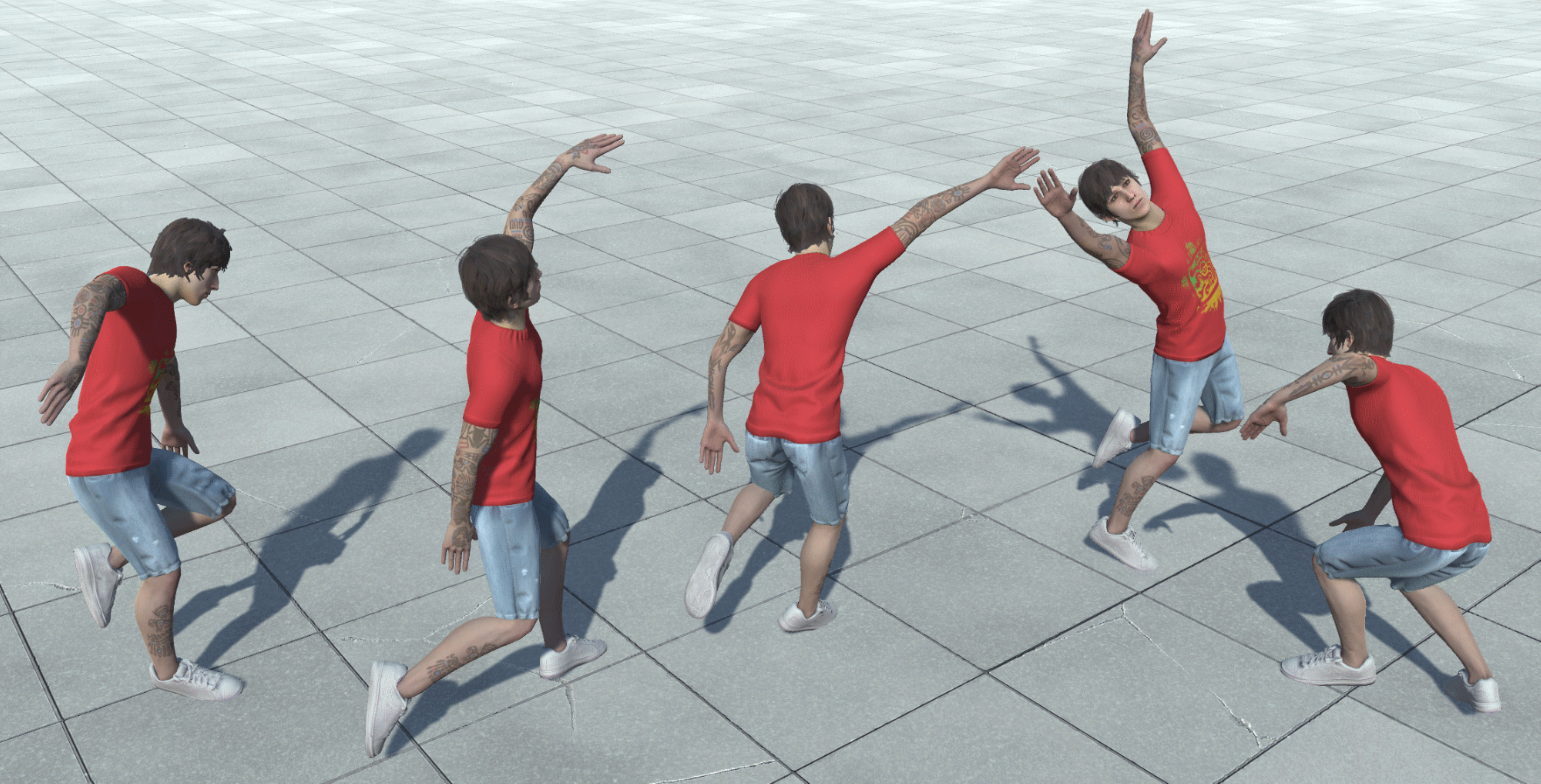}
        \label{fig:dance_male}
    }
  \caption{Synthesized motions for the characters by TPTN (from 0m38s to 0m45s and 2m19s to 3m0s in the accompanying video). Note that the transitions of continuous side somersaults and cartwheels are also \textit{not} in the training dataset.}
  \label{fig:flips_dances}
\end{figure*}

\noindent\textbf{Trajectory control loss}. For simplicity, we also represent this term as a Gaussian loss and integrate it to $L_{G}$. We only use the mean of the predicted PDF, $\hat{\mathbf{c}}_n^p$ and $\hat{\mathbf{c}}_n^d$, to generate motions. This term is helpful in interactive motion control when the user might occasionally input control signals. In this case, the predicted control signal values will be fed into the network to continue the motion synthesis.

\subsection{Training Details}
\label{sec:training_detail}
We train our TPTN using the AdamW optimizer ~\cite{loshchilov2017decoupled}. The initial learning rate is 1e-4 and will be decayed by multiplying it by 0.5 every 500 epochs. The batch size is set to be 56, and each sample in the batch is a motion clip of 300 consecutive frames denoted by $\mathbf{mc}$. Motion clips $\mathbf{mc}$ are obtained as follows: we repeatedly sample 300 frames for each motion sequence in the database from the first frame using a frame interval $mv_b=4$, i.e., the starting frame index, $f_{s+1}$, of the next 300 frames clip is equal to $f_{s}+mv_b$, where $f_s$ is the starting frame index of the current 300 frames.

We exploit the oversampling technique~\cite{ling1998data} to compensate for the imbalance of motion types in the dataset. We also increase the ratios of types that are hard to model for our TPTN. The oversampling is done in four steps: 1) Compute the ratio $r_e$ for all the motion types as the inverse of the number of motion types. In our case, $r_e$ equals $1/20$. 2) Compute the ratios $r_r$ for different types in the training data, which is done by dividing the number of motion frames for a motion type by the total number of motion frames. 3) Compute the new frame interval $mv$ as $mv = r_r / r_e * mv_b$, where $mv_b=4$ is the basic frame interval in our experiments. 4) Repeatedly sample 300 frames using the recomputed $mv$ to generate $\mathbf{mc}$ for training. Finally, we acquire a set of 60,962 $\mathbf{mc}$ samples for training, then randomly select 56 clips from the set in each iteration.

\noindent\textbf{Data augmentation}. To handle accumulated pose errors in the motion synthesis, we add additional independent identically distributed Gaussian noises to each sampled motion representation vector of training data to simulate the pose errors. The mean and standard deviation of the noise are selected to be $0$ and $0.05$.

\section{Experiments}
\label{sec:experiments}
We have implemented our algorithm using Pytorch 1.8.0 on a desktop PC with Intel(R) Xeon(R) E5-2678 CPU, 128 GB RAM, and one GeForce RTX 3090 24GB graphics card. Note that the performance we reported in the last paragraph of Section~\ref{sec:introduction} is computed by the reciprocal of the inference time tested on this desktop. Although the network can generate high-quality motions, slight foot slidings might still occur. If not mentioned, the IK algorithm is adopted to remove the foot slidings in generated motions according to the predicted foot contact labels.

\subsection{Metrics}
\label{sec:metrics}
In this paper, we utilize four metrics commonly used in recent research on human motion synthesis to evaluate the quality of generated motions quantitatively:

\noindent\textbf{Body movement}. Similar to~\cite{starke2020local}, we compute the body movement (BM) by the sum of the absolute angle updates of all the joints per frame. Given the joint angles $\mathbf{d}$, the BM can be computed as:
\begin{equation}
    BM=\frac{1}{T-1} \sum_{n=1}^{T-1} \sum_{j=0}^{22} \parallel \mathbf{d}_n^j - \mathbf{d}_{n-1}^j \parallel_1
    \label{eq:bm}
\end{equation}
This metric indicates how agile and various the generated motions appear. The smaller body movement indicates that the synthesized motions may be over-smoothed, while the bigger value means there may be more jitters in the generated motions.

\noindent\textbf{Average foot sliding}. The average foot sliding (AFS) is adopted from the metric proposed in~\cite{zhang2018mode} to estimate the amount of foot sliding in the generated motions. It can be computed as:
\begin{equation}
  AFS = \frac{1}{4(T-1)} \sum_{n=1}^{T-1} \sum_{j \in \{lf, lt, rf, rt\}} v_n^j(2-2^{\frac{h_n^j}{H^j}})
  \label{eq:afs}
\end{equation}
where $lf, lt, rf, rt$ represent the left/right foot/toe, $v_n^j$ is the velocity of the foot or toe on the XOZ plane, $h_n^j$ is its height from the plane, $H^j$ is a maximum threshold, which is the mean height for the foot (10.85cm) or toe (1.55cm) when they are in contact with the ground in our whole dataset, and the exponent is clamped between 0 and 1.

\noindent\textbf{Structural similarity index measure}.
The BM and AFS are all local metrics because their computations do not include the spatial and temporal correlations among different joints. Therefore, we adapt the structural similarity index measure (SSIM)~\cite{wang2004image} to measure the global similarity between generated motions and mocap data. We compute the SSIM of joint rotations represented by exponential maps over the temporal axis in sliding windows with sizes being 11 frames. We first subtract the minimum value of the corresponding mocap data sequence, computed using all joints' axis-angle values to ensure the computed SSIMs are non-negative. Then we slide the window one frame each time and average the SSIMs for all the windows as the final metric for the sequence. It can be formulated as:
\begin{equation}
  SSIM = \frac{(2\mu_1 \mu_2 + C_1)(\sigma_{12} + C_2)}{(\mu_1^2 + \mu_2^2 + C_1)(\sigma_1^2 + \sigma_2^2 + C_2)}
  \label{eq:ssim}
\end{equation}
where $\mu_1, \mu_2$ are the average values of the mean of the generated and the ground-truth sequence in sliding widows, $\sigma_1, \sigma_2$ are the corresponding standard variances and $\sigma_{12}$ is the covariance of two sequences. For the constants, $C_1=(K_1 L)^2$ and $C_1=(K_2 L)^2$, where $L$ is computed as the difference between the maximum and minimum in the ground-truth sequence, $K_1=0.01$ and $K_2=0.03$ by default. SSIM will be close to 1 if two motion sequences are spatially and temporally similar.

\noindent\textbf{Mean per joint position error}. Given the joint positions $\mathbf{p}$ that are relative to the root joint position, the mean per joint position error (MPJPE) in~\cite{ionescu2013human3} can be formulated as:
\begin{equation}
    MPJPE=\frac{1}{23T} \sum_{n=0}^{T-1} \sum_{j=0}^{22} \parallel \hat{\mathbf{p}}_n^j - \mathbf{p}_{n}^j \parallel_2
    \label{eq:mpjpe}
\end{equation}
where $\hat{\mathbf{p}}_n^j, \mathbf{p}_{n}^j$ are the $n_{th}$ frame's $j_{th}$ joint's generated and ground-truth positions, respectively.

Since the motions are generated according to constraints defined by the user for controllable motion synthesis, it is usually impractical to compute the difference between synthesized motions and their corresponding ground-truth motions. Therefore, MPJPE can not be directly applied to evaluate the performance of generated motions. We manage to compute MPJPE averaged over test sequences of different motion types since we can extract the control signals from the ground-truth mocap data of test sequences to guide the TPTN to generate motions similar to the ground truth.

\begin{figure}[t]
  \centering
    \includegraphics[width=0.95\linewidth]{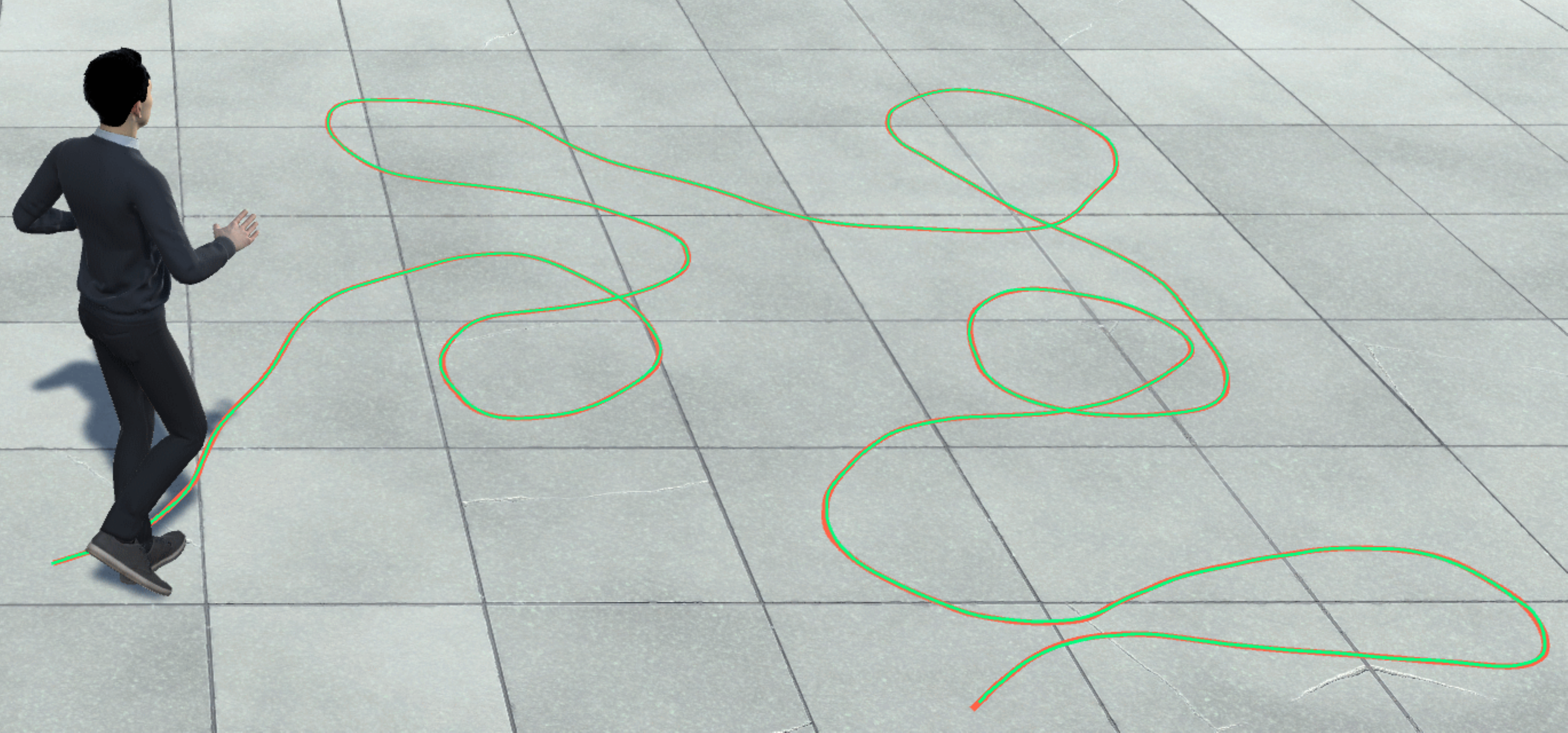}
  \caption{Trajectory-following result synthesized by TPTN following a trajectory with large curvatures (from 1m57s to 2m15s in the accompanying video). The distance between the synthesized trajectory (the green line) and the input trajectory (the orange line) is 0.513cm/frame.}
  \label{fig:traj_following}
\end{figure}

\begin{figure}[t]
\vspace{-1.0em}
  \centering
    \subfloat[]{
       \includegraphics[width=0.47\linewidth]{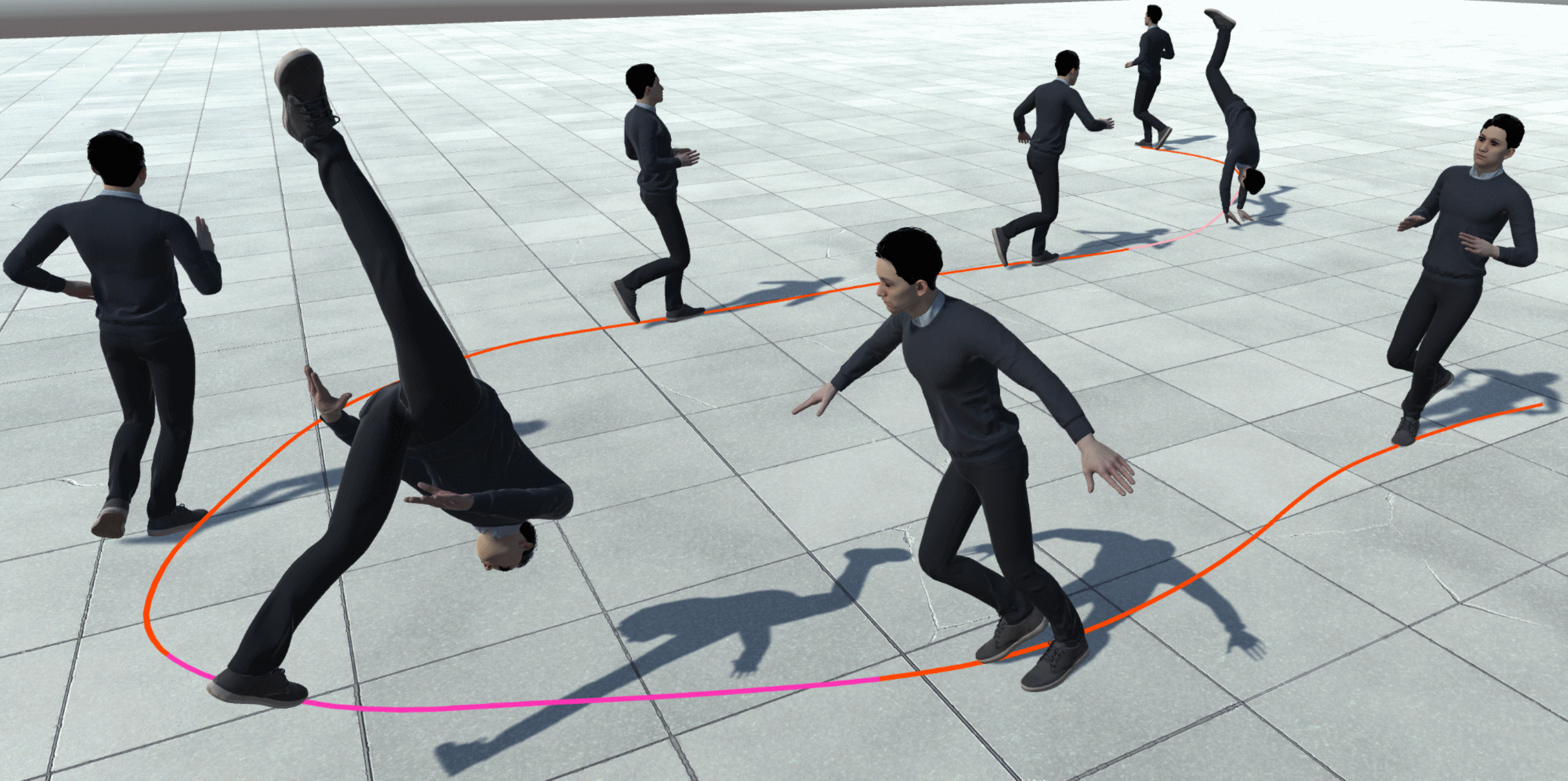}
       \label{fig:same_traj_type1}
    }
    \subfloat[]{
        \includegraphics[width=0.47\linewidth]{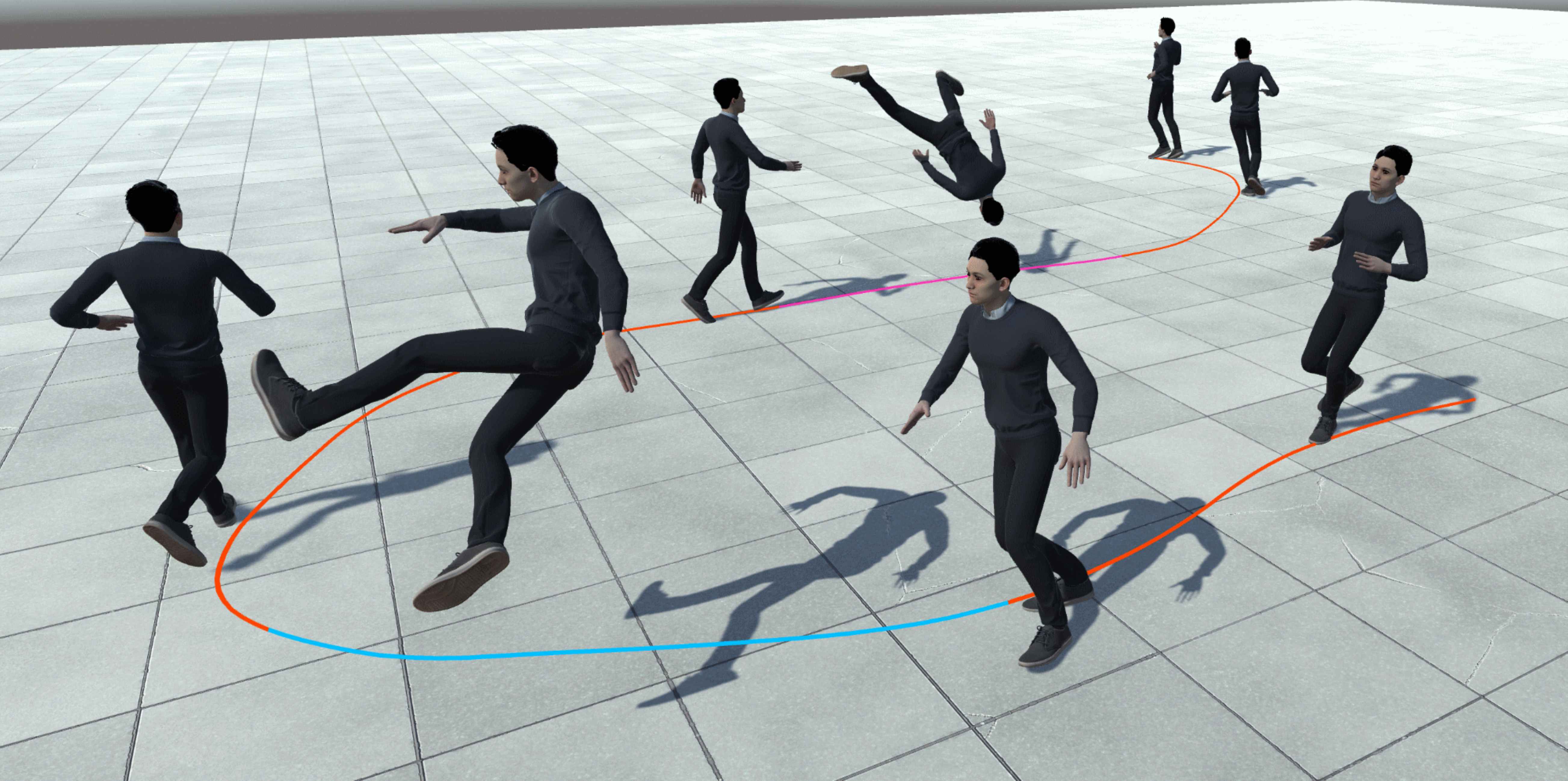}
        \label{fig:same_traj_type2}
    }
  \caption{Trajectory-following results (from 1m33s to 1m44s in the accompanying video). The trajectories in the two pictures are the same but assigned with different motion type information. (a) A synthesized motion transitioning from running to side somersault, then to cartwheel, and ending with running. (b) A synthesized motion transitioning from running to whirlwind kicking-2, then to side somersault and ending with running. We use different colors to represent different motion types at different parts of trajectories. Details can be found in the supplementary material.}
  \label{fig:same_traj_different_types}
\end{figure}

\begin{figure}[t]
  \centering
    \includegraphics[width=0.95\linewidth]{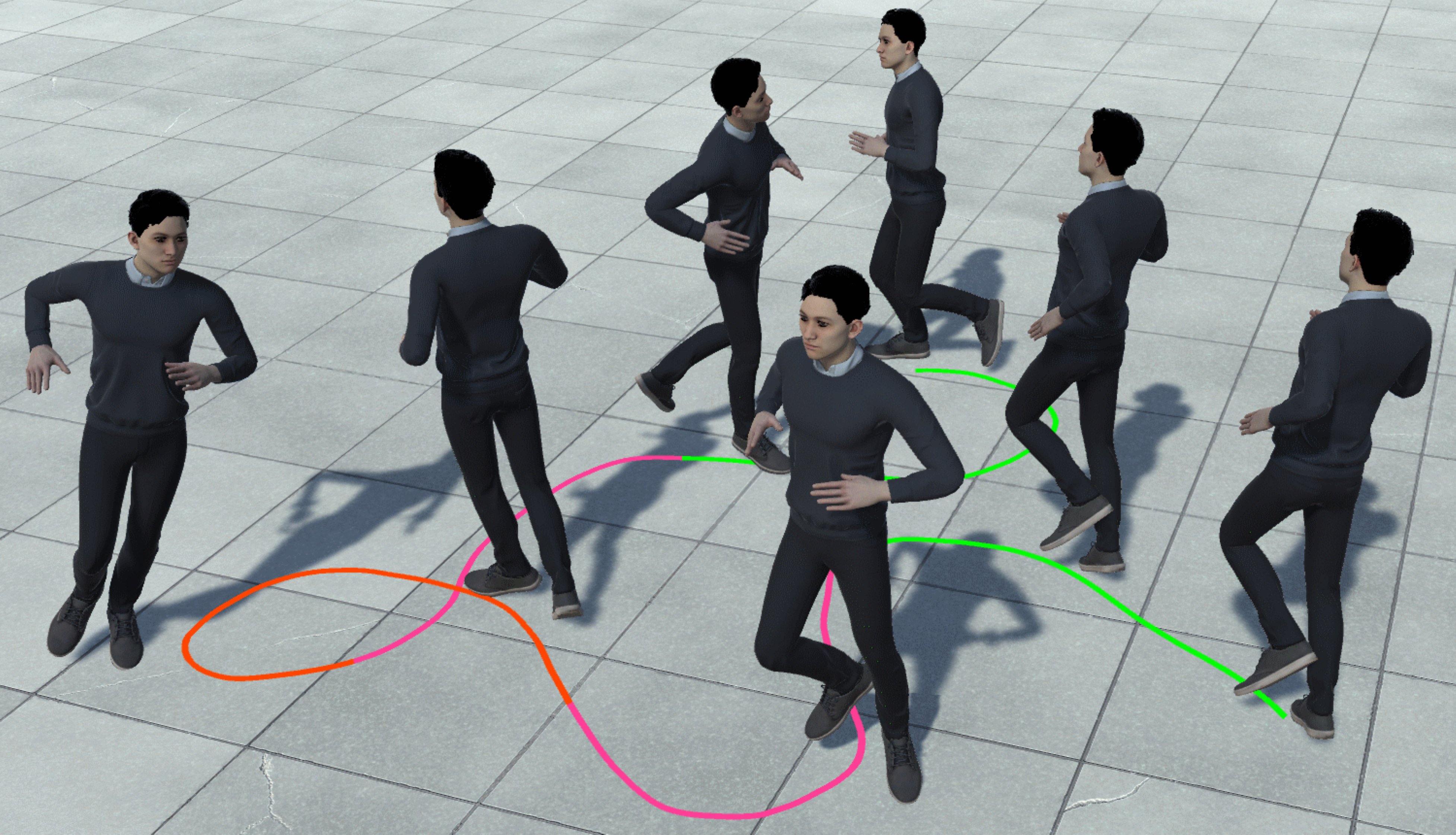}
  \caption{Our TPTN can synthesize motions heading along a complex trajectory transitioning from jumping-R to jumping-L, to running, then to jumping-L, and ending with jumping-R (from 1m21s to 1m32s in the accompanying video).}
  \label{fig:complex_traj}
\end{figure}

\begin{figure}[t]
\vspace{-1.0em}
  \centering
    \subfloat[]{
       \includegraphics[width=0.47\linewidth]{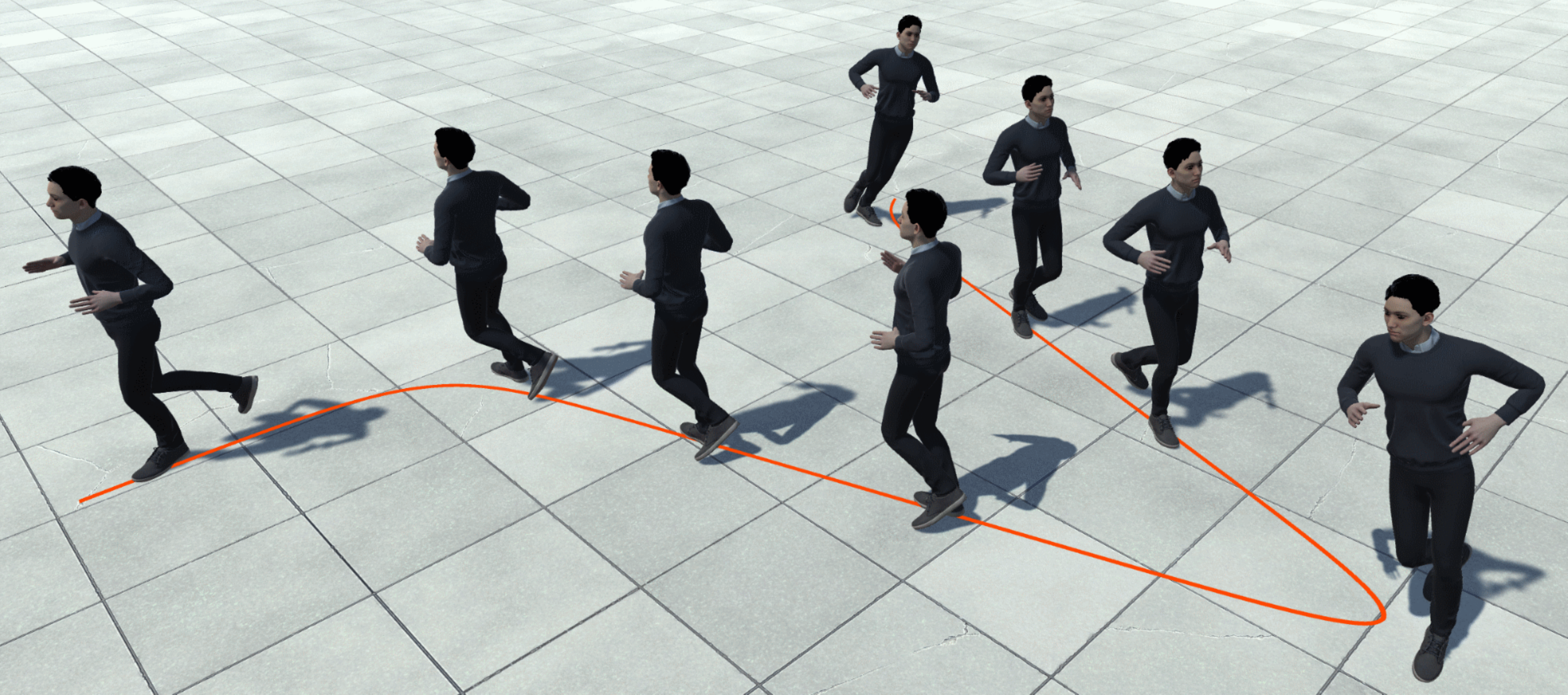}
       \label{fig:star_corner}
    }
    \subfloat[]{
        \includegraphics[width=0.47\linewidth]{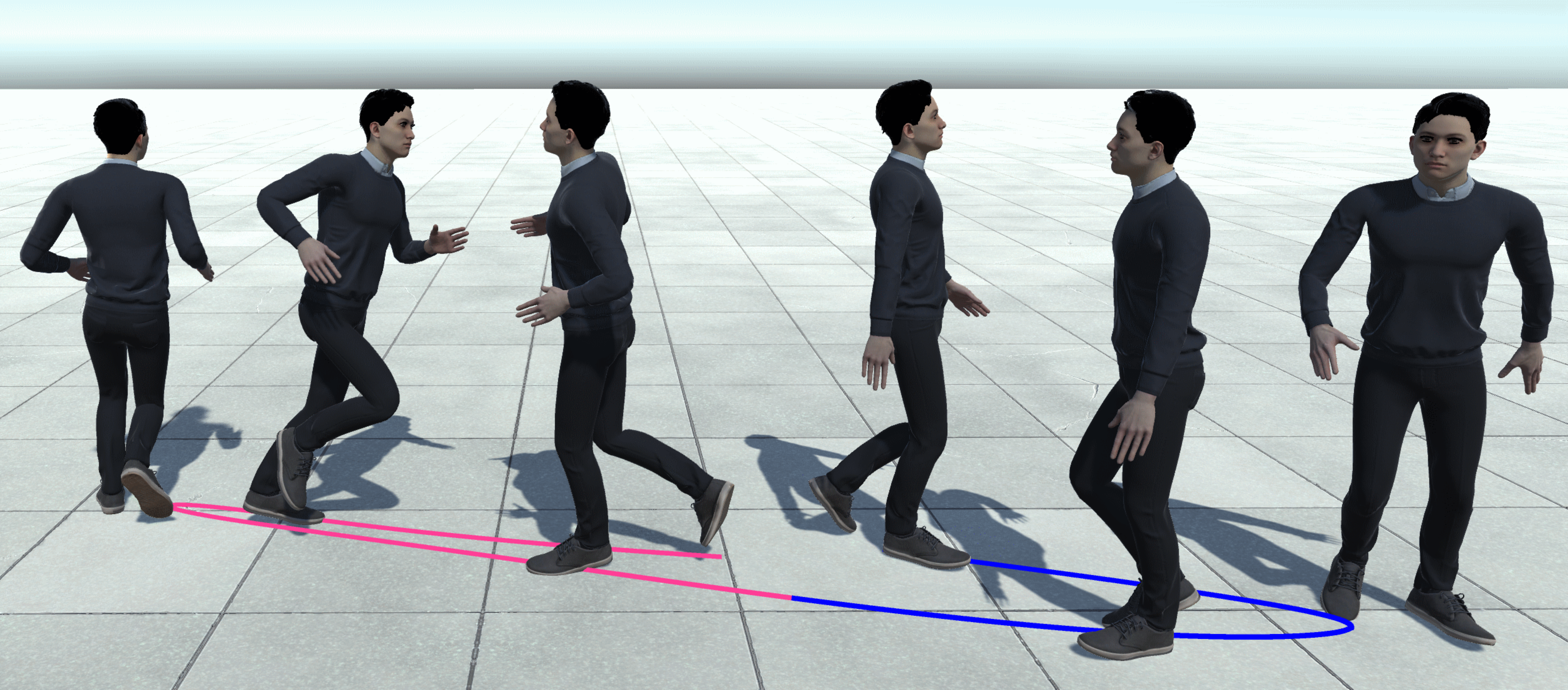}
        \label{fig:uturn}
    }
  \caption{TPTN can synthesize sharp turns (from 1m44s to 1m57s in the accompanying video). (a) Running along a pentagram corner. (b) A U-turn walking and jumping-L. IK is disabled in this experiment.}
  \label{fig:sharp_turns}
\end{figure}

\subsection{Motion Synthesis Results}
\label{sec:motion_results}
\noindent\textbf{Synthesizing different types of motions}.
In this experiment, we use the first 30 frames of each sequence in the test dataset as initial frames and input them to the TPTN. The control signals $\mathbf{c}_n^s$ are extracted from the sequences in the test dataset. The motion type information $\mathbf{c}_n^t$ is specified by the user, while $\mathbf{c}_n^p$ and $\mathbf{c}_n^d$ are predicted by the network. The control signals for initial frames are automatically computed according to the motion data. Examples in Fig.~\ref{fig:teaser}\&~\ref{fig:flips} illustrate that the TPTN can generate natural and realistic heterogeneous motions, such as running and cartwheels. While the motion transitions in these two sequences between different types of motions do not exist in our training dataset, the TPTN still successfully generates smooth transitions, demonstrating our network's generalization ability. Furthermore, TPTN can also generate dances for the two dancers, as shown in Fig.~\ref{fig:dance_female}\&~\ref{fig:dance_male}.

\noindent\textbf{Following user-specified trajectories}. It is desirable to synthesize different types of motions along a specified trajectory in many applications, for instance, motion planning in video games. We allow users to specify a motion trajectory $\mathcal{J}$ on the XOZ plane with additional motion types and velocities and then map the trajectory information into the control signals $\mathbf{c}_n^p$, $\mathbf{c}_n^d$ and $\mathbf{c}_n^t$. The trajectory $\mathcal{J}$ is defined as $\mathcal{J}=\{\{\mathbf{J}_i,\mathbf{t}_i,v_i\}, i=1,..,k\}$, where $\mathbf{J}_i$ is the $i_{th}$ part in the trajectory represented as densely sampled 2D points, $\mathbf{t}_i$ and $v_i$ are the motion types and a scalar velocity value associated to it. For two adjacent parts of the trajectory with different motion types, we set up 40 transitional frames and interpolate motion type signals for these frames (see the supplementary material for the details of the representation and computation of $\mathcal{J}$).

Given some initial frames extracted from the test dataset and user-specified trajectory $\mathcal{J}$, TPTN can synthesize motions that accurately follow the user-specified trajectory as pictured in Fig.~\ref{fig:traj_following}. We compute the trajectory distance by averaging the closest distance between a projected root position and a target trajectory in each frame. The distance between the synthesized trajectory (the green line) and the input one (the orange line) in Fig.~\ref{fig:traj_following} is 0.513cm/frame. TPTN can generate motions with different types while following the same trajectories (Fig.~\ref{fig:same_traj_different_types}), even for a trajectory with large curvatures and frequent change of motion types (Fig.~\ref{fig:complex_traj}). In all the figures, different trajectory colors indicate different types of motions. Additionally, the trajectory distances of three examples in Fig.~\ref{fig:same_traj_different_types} \& Fig.~\ref{fig:complex_traj} are 1.988cm/frame, 1.749cm/frame, and 0.454cm/frame respectively.

It is harder for a person to control their body trajectory in the air than being on the ground, and so is our TPTN. For example, the character is frequently in the air to perform the cartwheel or whirlwind kicking motion, as shown in Fig.~\ref{fig:same_traj_different_types}. The trajectory distances of this example are larger than those examples shown in Fig.~\ref{fig:traj_following} and Fig.~\ref{fig:complex_traj}.

We further test TPTN's ability to synthesize motions along trajectories with sharp turns. As shown in Fig.~\ref{fig:star_corner}, the character turns around 144 degrees along a pentagram corner and turns nearly 180 degrees in Fig.~\ref{fig:uturn}. It demonstrates that TPTN can learn to generate turn motions in different angles from our dataset with sharp turn motions.

\noindent\textbf{Interactive control}. We implement our interactive control demo on a desktop PC with Intel(R) Core(TM) i7-7700K CPU, 16GB RAM, and one GeForce GTX 1080 Ti 12GB graphics card in the Unity3D engine, and the neural network is queried through a TCP socket interface to input the character motions into the engine. Our interactive demo allows the user to control the motion types, forward directions, and velocities through a keyboard (refer to the supplementary material for details). We generate ten frames using the network before sending them to Unity3D for display, resulting in a 0.17s delay or so. Besides, the response time of our network to the user input is around 0.5s. Therefore, the responsiveness introduced in~\cite{starke2020local} is around 0.7s for TPTN. For comparisons, the average responsiveness is about 0.9s for PFNN~\cite{holden2017phase}, 1.1s for MANN~\cite{zhang2018mode}, and 0.7s for the method in~\cite{starke2020local}. Note that these values are obtained by converting the bars back to the numbers through meticulous measurements from Fig.14 in~\cite{starke2020local}. This demo verifies that our method can respond to user input fast. Please see the supplementary material and accompanying video from 0m45s to 1m17s for the demo.

\begin{table}[t]
  \centering
  \resizebox{0.4\textwidth}{!}{
    \begin{tabular}{ccccc}
        \toprule[0.8pt]
        Metrics        & GT     & MVAE      & HTSS    & TPTN             \\ \midrule
        BM (deg/frame) & 82.740 & 101.353   & 137.601 & \textbf{78.401}  \\ \midrule
        AFS (cm/frame) & 0.208  & 0.534     & 0.529   & \textbf{0.331}   \\ \midrule
        SSIM           & 1.0      & 0.977     & 0.979   & \textbf{0.985}   \\ \midrule
        MPJPE        & 0.0      & 2.905     & \textbf{1.771}   & 1.906      \\ \bottomrule[0.8pt]
    \end{tabular}
  }
  \caption{Comparisons on the metrics for all test motions containing dances. MVAE and HTSS are trained on the same dataset as TPTN. GT: ground-truth mocap data. IK is disabled in this experiment. Please refer to the supplementary material for these metrics on different types of motions.}
  \label{tab:comps_dance}
\end{table}

\begin{table}[t]
  \centering
  \resizebox{0.48\textwidth}{!}{
    \begin{tabular}{cccccc}
        \toprule[1pt]
        Metrics        & GT     & MVAE      & HTSS    & NeuralLayering & TPTN             \\ \midrule
        BM (deg/frame) & 95.752 & 86.477    & 134.040 & 112.072        & \textbf{89.274}  \\ \midrule
        AFS (cm/frame) & 0.234  & 0.480     & 0.363   & 0.351          & \textbf{0.302}   \\ \midrule
        SSIM           & 1.0    & 0.985     & 0.990   & \textbf{0.994} & 0.992  \\ \midrule
        MPJPE        & 0.0    & 3.809     & \textbf{1.831}   & 2.185 & 1.885  \\ \bottomrule[1pt]
    \end{tabular}
  }
  \caption{Comparisons on the metrics for non-dance motions. MVAE, HTSS, and NeuralLayering are trained on the dataset without dances. GT: ground-truth mocap data. IK is disabled in this experiment. Please refer to the supplementary material for these metrics on different types of motions.}
  \label{tab:comps}
\end{table}

\begin{figure}[t]
  \vspace{-0.8em}
  \centering
    \subfloat[MVAE]{
        \includegraphics[width=0.46\linewidth]{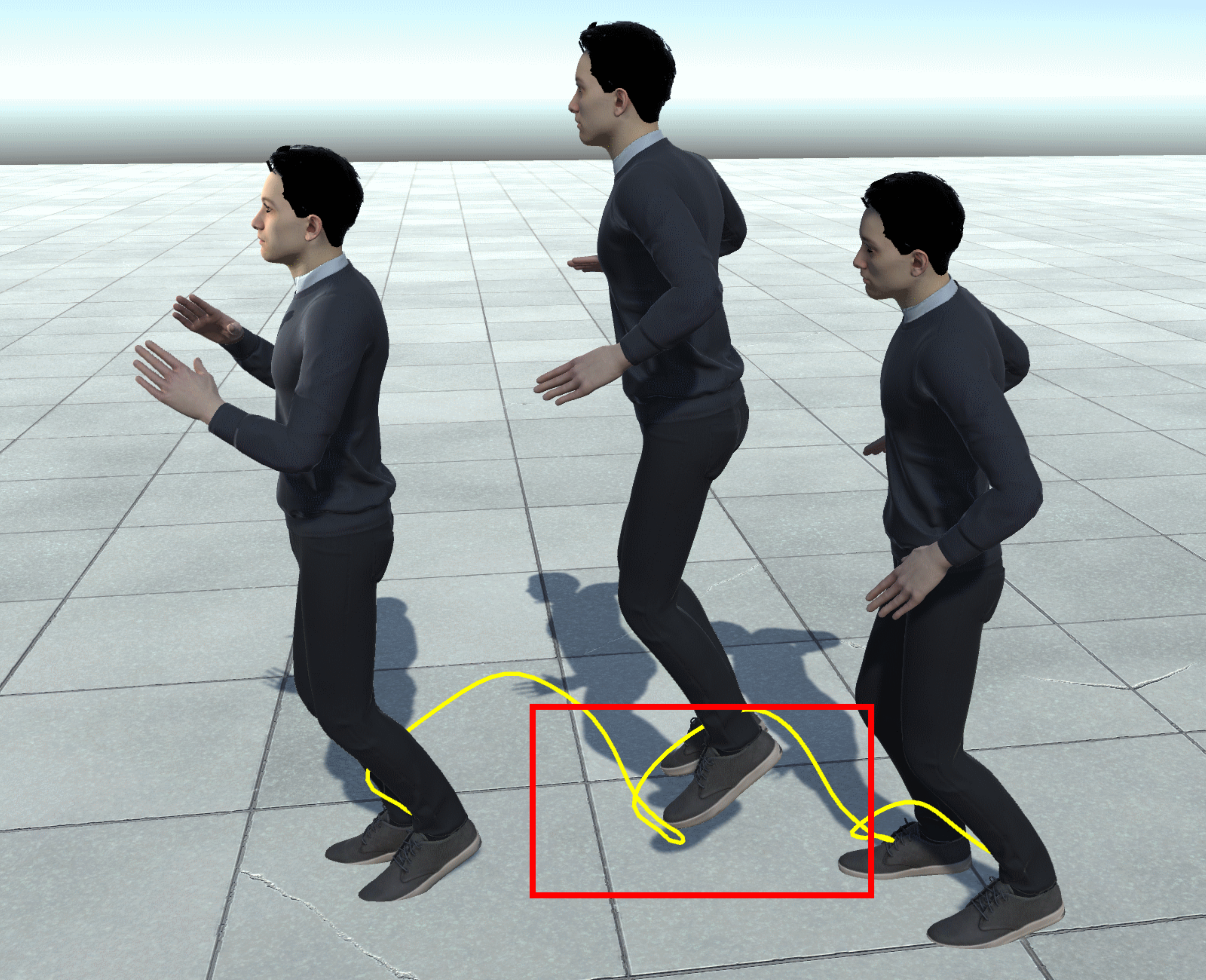}
        \label{fig:mvae_sliding}
    }
    \subfloat[HTSS]{
        \includegraphics[width=0.46\linewidth]{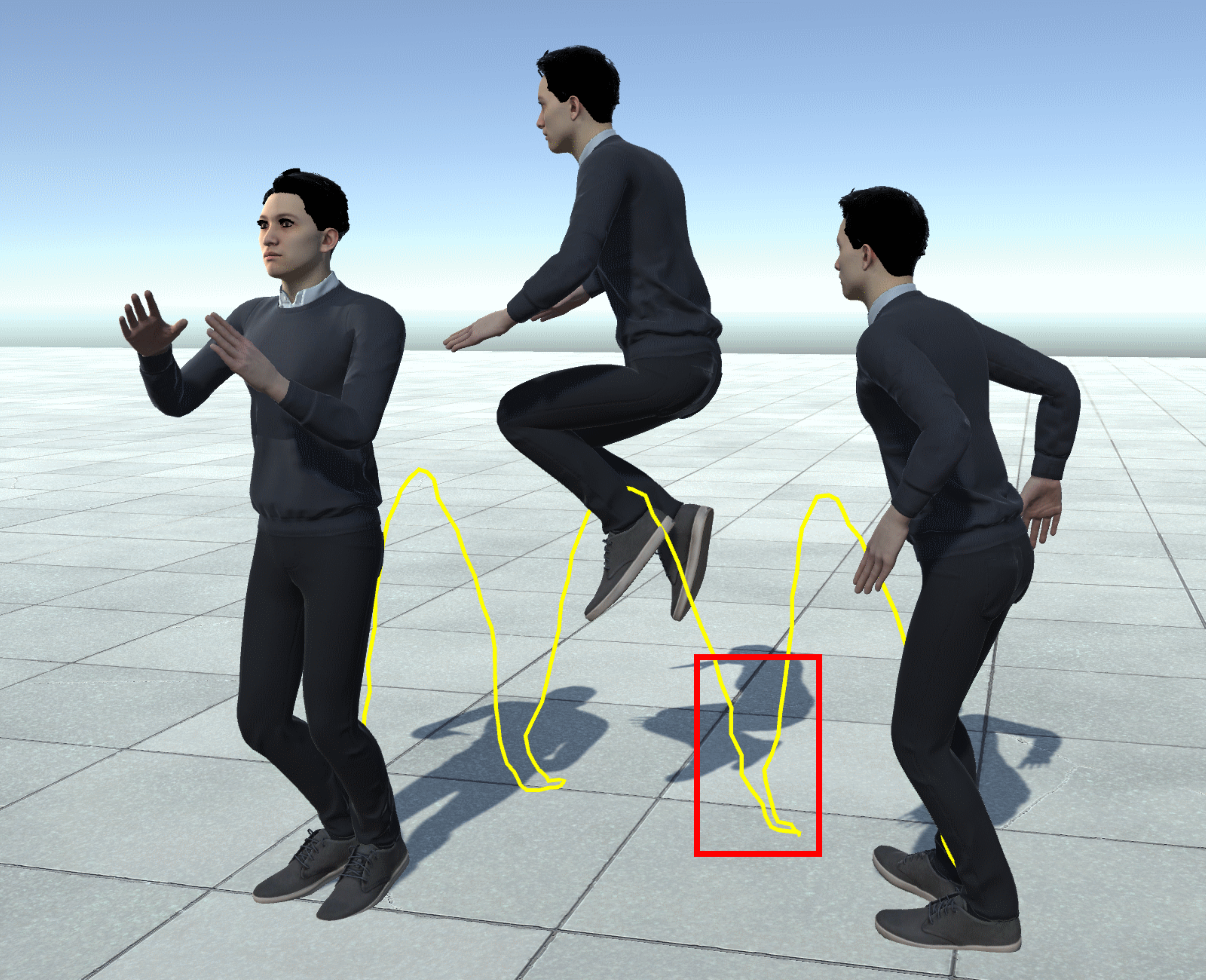}
        \label{fig:htss_sliding}
    }
    \quad
    \subfloat[NeuralLayering]{
        \includegraphics[width=0.46\linewidth]{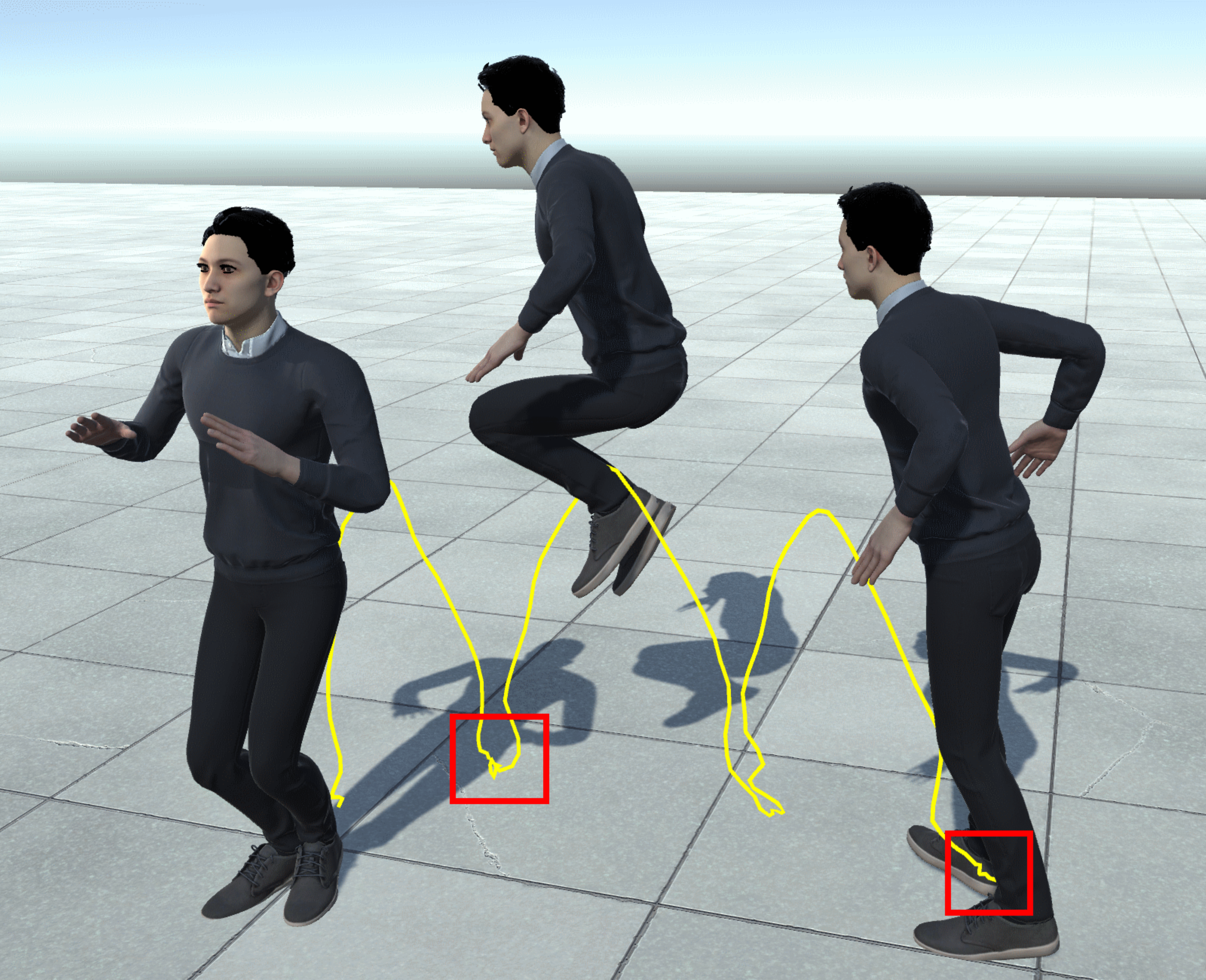}
        \label{fig:neural_sliding}
    }
    \subfloat[TPTN]{
       \includegraphics[width=0.46\linewidth]{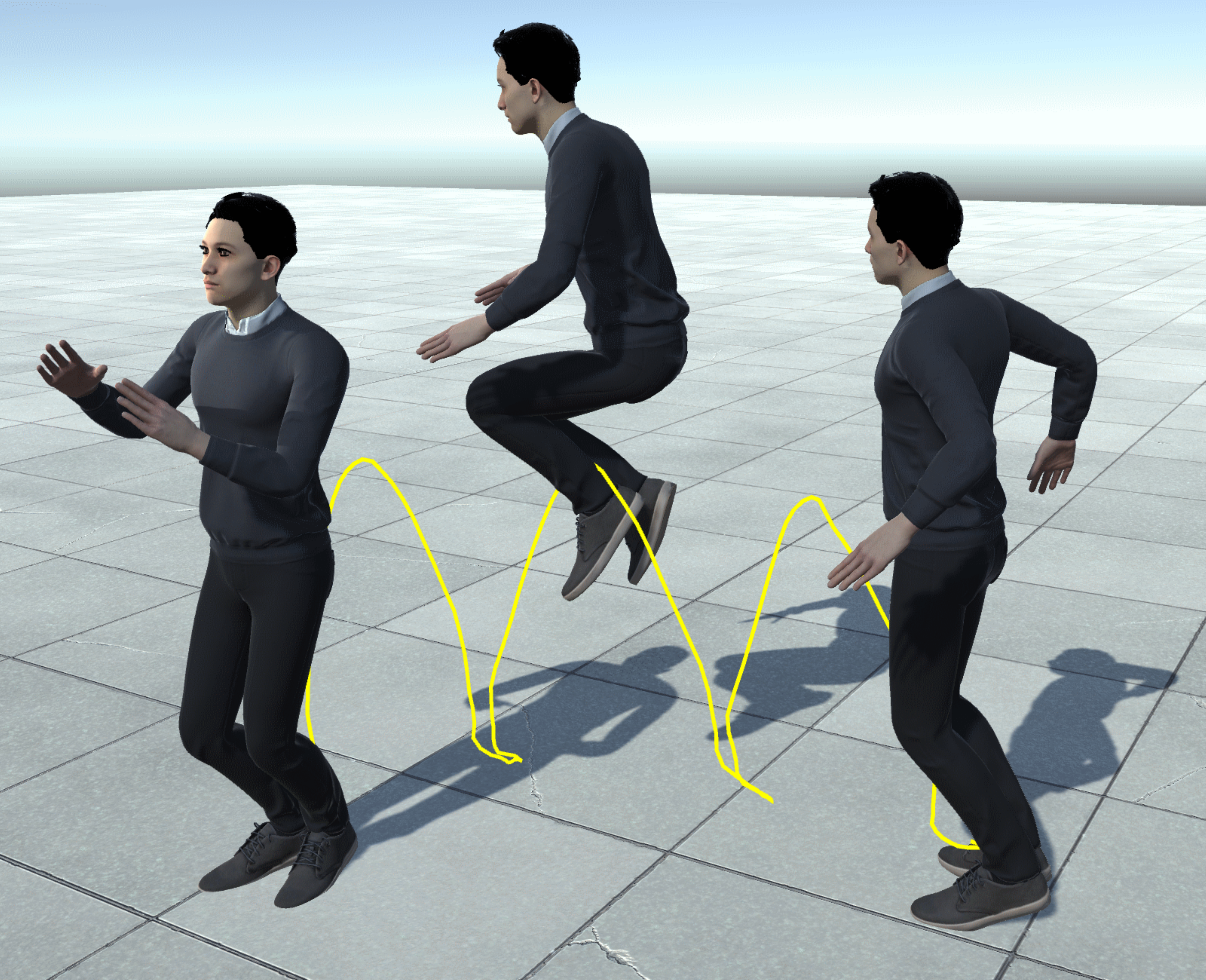}
       \label{fig:tptn_sliding}
    }
  \caption{The left foot's trajectories of jumping-B motions generated by (a) MVAE, (b) HTSS, (c) NeuralLayering, and (d) TPTN. As indicated by the red rectangles, the motion generated by MVAE has an apparent lower height, the motion generated by HTSS has wrong feet poses and high-frequency jitters, while the primary artifacts of motion generated by NeuralLayering are high-frequency jitters. In contrast, our TPTN can generate a smoother motion sequence than baseline networks. IK is disabled in this experiment.}
  \label{fig:slidings}
\end{figure}

\subsection{Comparisons}
\label{sec:comparison}
\subsubsection{Comparisons with Models Trained on Our Dataset} We compare our TPTN model with state-of-the-art baseline models on the quality of generated motions measured by the metrics in Section~\ref{sec:metrics}. The metric statistics computed in the comparisons are shown in Tab.~\ref{tab:comps_dance}\&Tab.\ref{tab:comps}.

\label{comp_our_dataset}
\noindent\textbf{Baseline Models}: 

\begin{itemize}
    \item MVAE: the auto-regressive conditional VAE used for reinforcement learning to produce desired goal-directed locomotion movement in~\cite{ling2020character}.
    \item HTSS: the hierarchical two-stream sequential model generating pose sequences from input textural descriptions in~\cite{ghosh2021synthesis}.
    \item NeuralLayering: the neural network imitating animation layering for synthesizing martial art movements in~\cite{starke2021neural}.
\end{itemize}
We make some adaptations to the baseline models to better test their ability to model heterogeneous motions. To adapt MVAE such that it can be applied to control various motion types in our dataset, we first train VAE by incorporating the motion type to the input of the encoder, decoder and the control policy network, as discussed in the MVAE paper. However, experimental results show that the generated motions' quality is not as good as the authors reported after training. The reason might be the simple kinematics regularizer used in the reinforcement learning algorithm in this method, which is not as direct as the reward function in~\cite{peng2018deepmimic} that takes the pose similarity into consideration. Another reason might be that the heterogeneous motions are more difficult to model than locomotion, like walking or running. Therefore, we obviate the policy and controller networks but directly concatenate control signals to the past pose that is input to the encoder and decoder; also, we remove the current pose from the input to break the gap between training and inference. For HTSS, we first map the concatenation of control signals to features by a linear layer, then add them with motions' features both in the encoder and decoder to keep the dimension of the first layer unchanged. We only use the locomotion control module for NeuralLayering because our dataset has no multi-character interactions. As a result, their parameters are 2.03M, 4.7M and 25.21M, respectively. Note that Starke et al.~\cite{starke2021neural} have not released their codes; thus, we re-implement their algorithms according to their paper and appendix. TPTN, MVAE, and NeuralLayering are trained for 2000 epochs, and HTSS for 300 epochs with an initial learning rate being 5e-4. Except for the modifications mentioned earlier, the rest hyper-parameters and settings for baseline networks remain the same as reported in their papers.

\noindent\textbf{Metric Computation}:The metrics introduced in Section~\ref{sec:metrics} are used in comparisons. We thus leverage the motions in the test dataset, 12 non-dance motions and two dances, as ground truth and extract control signals from the ground-truth motions as described in Section~\ref{sec:control_signal}. Motion sequences are then synthesized by the trained models using the first 87 frames of each sequence as initial frames with extracted control signals. All the generated motions will be input to the metric computation together with ground truth motions.

\begin{figure}[t]
  \vspace{-0.8em}
  \centering
    \subfloat[MVAE]{
        \includegraphics[width=0.23\linewidth]{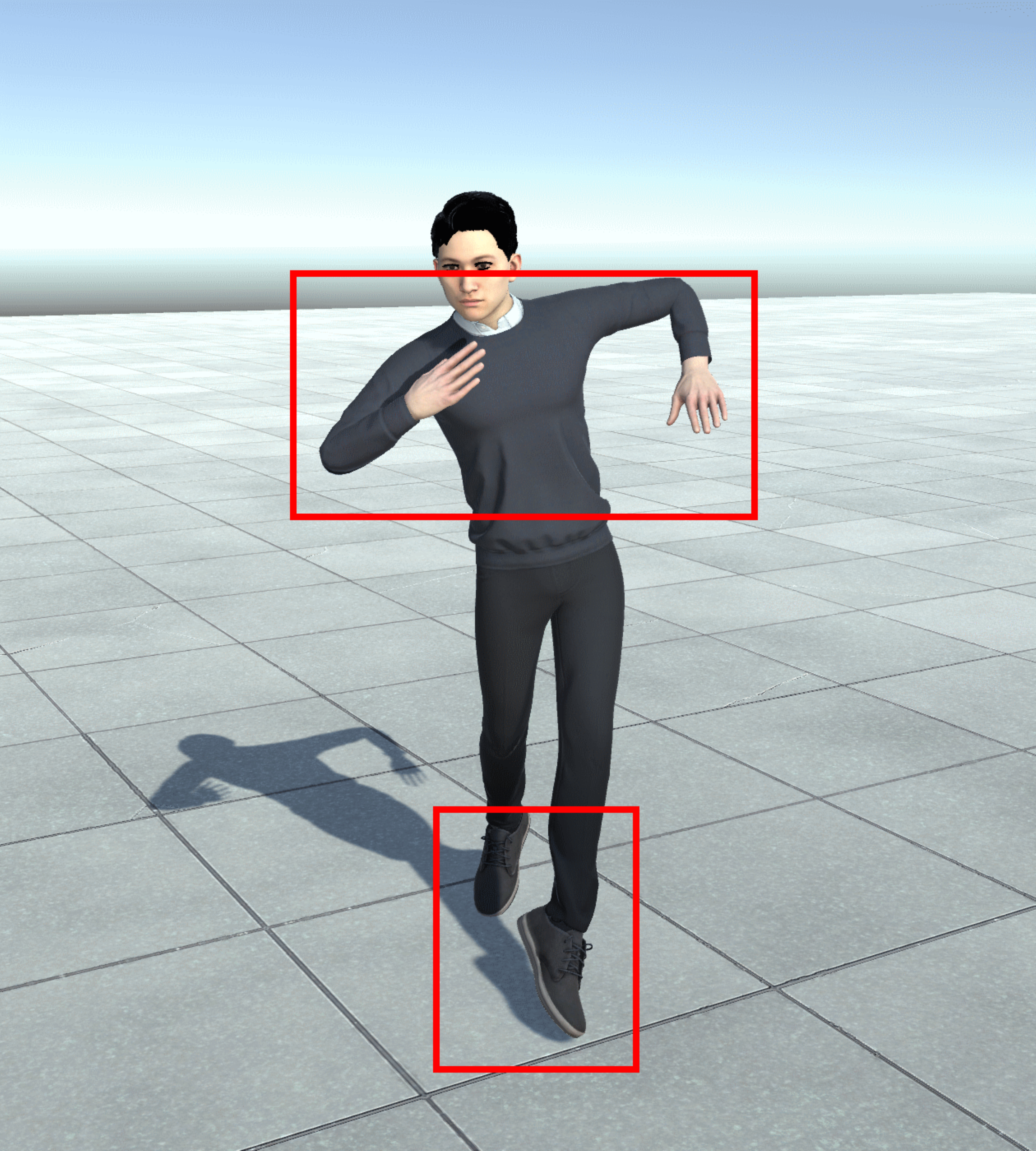}
        \label{fig:mvae_us}
    }
    \subfloat[HTSS]{
        \includegraphics[width=0.23\linewidth]{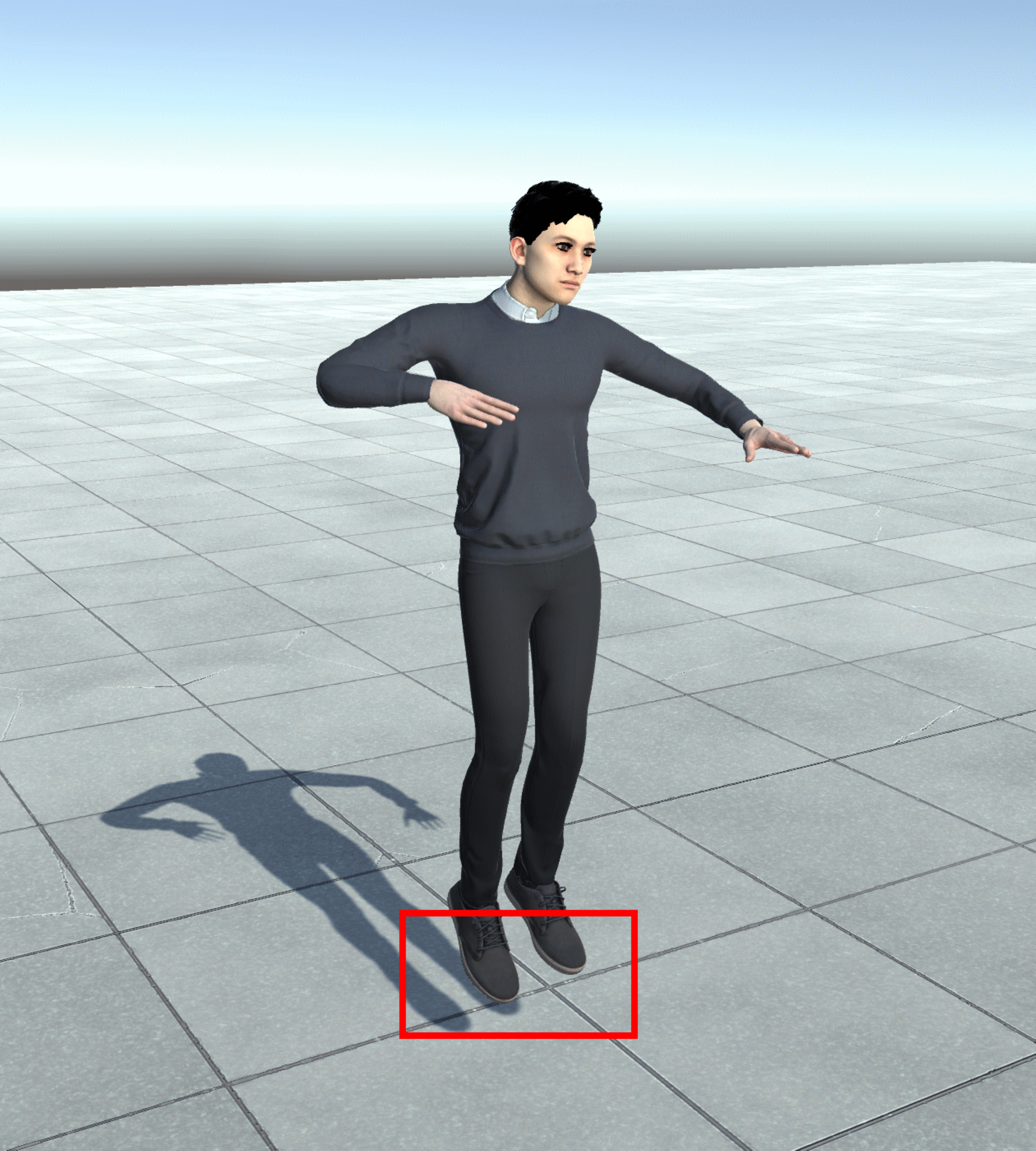}
        \label{fig:htss_us}
    }
    \subfloat[NL]{
        \includegraphics[width=0.23\linewidth]{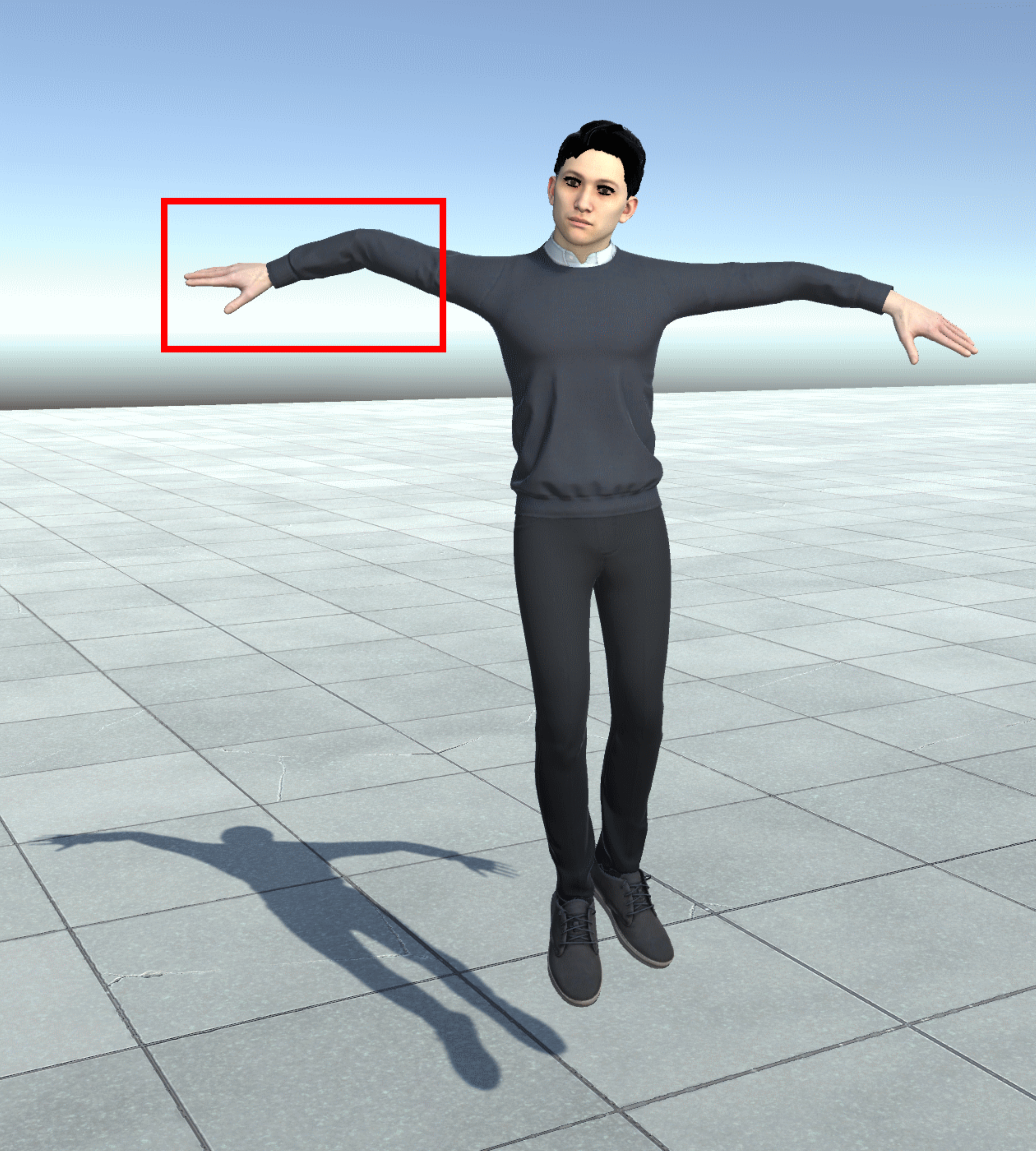}
        \label{fig:neural_us}
    }
    \subfloat[TPTN]{
       \includegraphics[width=0.23\linewidth]{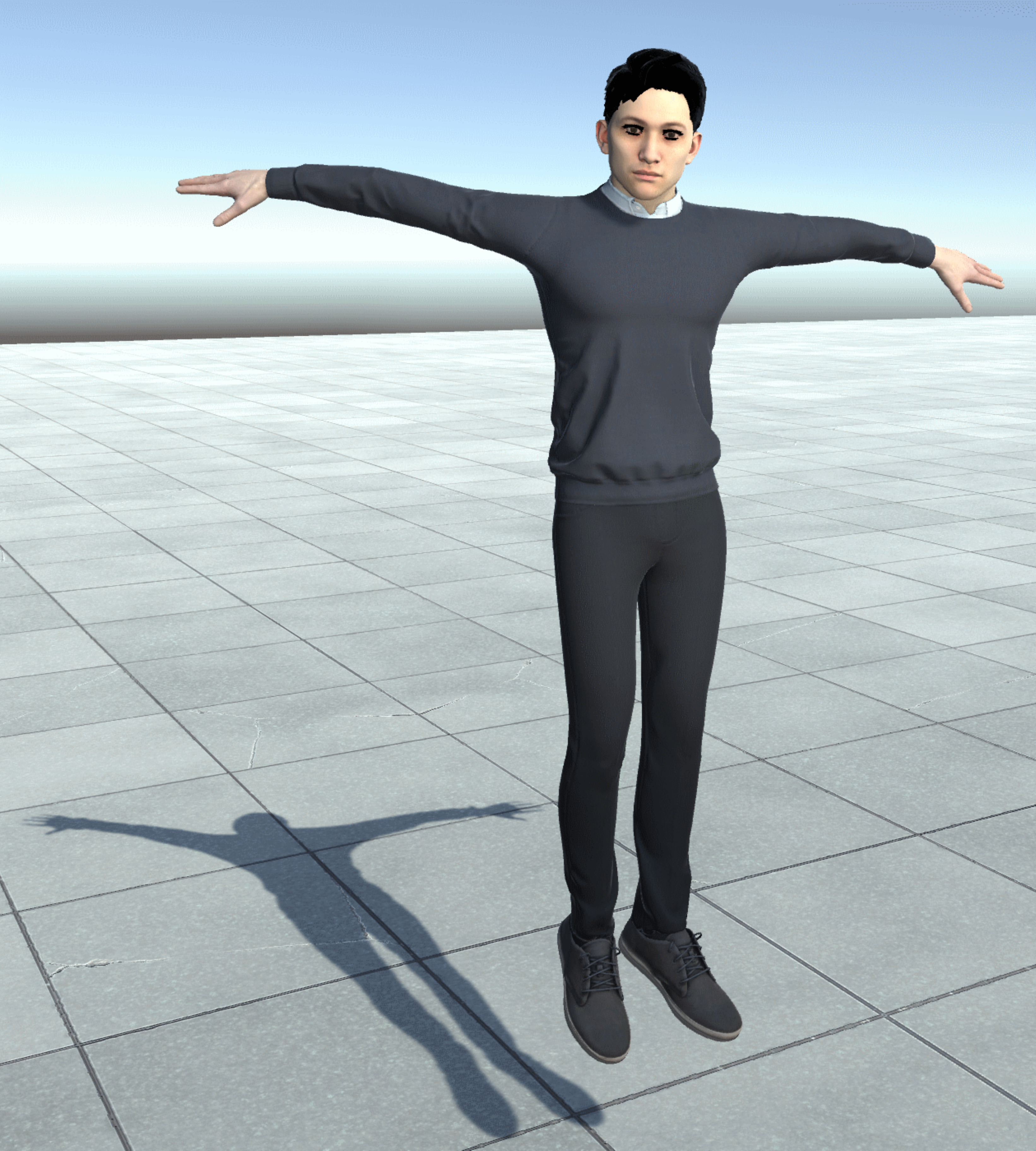}
       \label{fig:tptn_us}
    }
  \caption{The $273_{th}$ frame of motions transitioning from running to twist generated by (a) MVAE, (b) HTSS, (c) NeuralLayering, and (d) TPTN. NL is short for NeuralLayering. As pictured by the red rectangles, when the character is twisting, the arms and feet are more natural in the motion generated by TPTN than those generated by MVAE, HTSS and NeuralLayering. IK is disabled in this experiment.}
  \label{fig:user_study}
\end{figure}

\noindent\textbf{Training Strategies and Statistics}: We first used the same dataset as TPTN to train MVAE, HTSS, and NeuralLayering for fair comparisons. Since the code of the NeuralLayering method is not available, we implement this method by ourselves. However, we found that NeuralLayering performed not well on dancing motions with our implementation. Since our implementation of NeuralLayering might need to be fine-tuned to achieve the best performance, to avoid the superficial conclusion, we thus only compare TPTN with MVAE and HTSS using all the generated test motions that contain dances and report the metrics in Tab.~\ref{tab:comps_dance}. Moreover, we re-trained NeuralLayering from scratch on our dataset without dancing motions and obtained far better results. To compare with baseline models on our dataset without dancing motions, we also re-trained MVAE and HTSS on the same dataset without dances and computed the metrics using the generated test non-dance motions for all the models in Tab.~\ref{tab:comps}. \emph{Note that our TPTN is only trained with our full dataset in all the comparisons in Tab.~\ref{tab:comps_dance}\&Tab.~\ref{tab:comps}.}

As shown in Tab.~\ref{tab:comps_dance}\&Tab.~\ref{tab:comps}, TPTN outperforms the baseline models on nearly all metrics and achieves the second-best score on SSIM and MPJPE when comparing non-dance motions. The larger values of BM and AFS for HTSS indicate that there might be jitters in the motions generated by HTSS, as depicted in Fig.~\ref{fig:htss_sliding}. Because HTSS is a designed sequence-to-sequence model for text-to-motion translation, the big discrepancy of control signals might be the reason that it cannot generate high-quality motions; while its part-based design might still help it synthesize heterogeneous motions like dances as indicated in Tab.~\ref{tab:comps_dance}. Furthermore, the HTSS inference speed is 12fps, slower than our network. We can also see from Fig.~\ref{fig:slidings} that the motion generated by TPTN is smoother, and the motion generated by MVAE is over-smoothed because the jumping height is apparently lower than the mocap data. Fig.~\ref{fig:user_study} illustrates that TPTN can generate motions more similar to the mocap data; we thus deem that they are more natural. Please refer to the accompanying video from 3m08s to 4m21s for these results for better visual quality.

A common problem with TPTN, MVAE, and HTSS is that the dance quality generated for the female dancer is lower. The reason is that her skeleton size is significantly smaller than the other two characters, and her motions are the least (35,615 frames for training, while 45,005 for the male dancer and 92,336 for the gymnast).

\subsubsection{Comparisons on PFNN Dataset}
To verify our TPTN's ability to generate locomotion, we trained our model on the dataset released by PFNN~\cite{holden2017phase} and then compared TPTN's performance with PFNN. Since PFNN did not provide a test dataset, we thus randomly selected five sequences from the PFNN dataset as our test dataset, which are not included in the training. Moreover, PFNN fitted the terrain for each frame separately, which may result in noncontinuous terrain for the sequence, thus is inappropriate for a sequence-based model like TPTN. Hence, we adopt the fitting method in~\cite{holden2020learned}. In addition, we process the motion data and control signals as described in Section~\ref{sec:data_process} and use the same settings described in Section~\ref{sec:training_detail} to train our TPTN on this dataset.

\begin{figure}[t]
\vspace{-1.0em}
  \centering
    \subfloat[PFNN]{
       \includegraphics[width=0.47\linewidth]{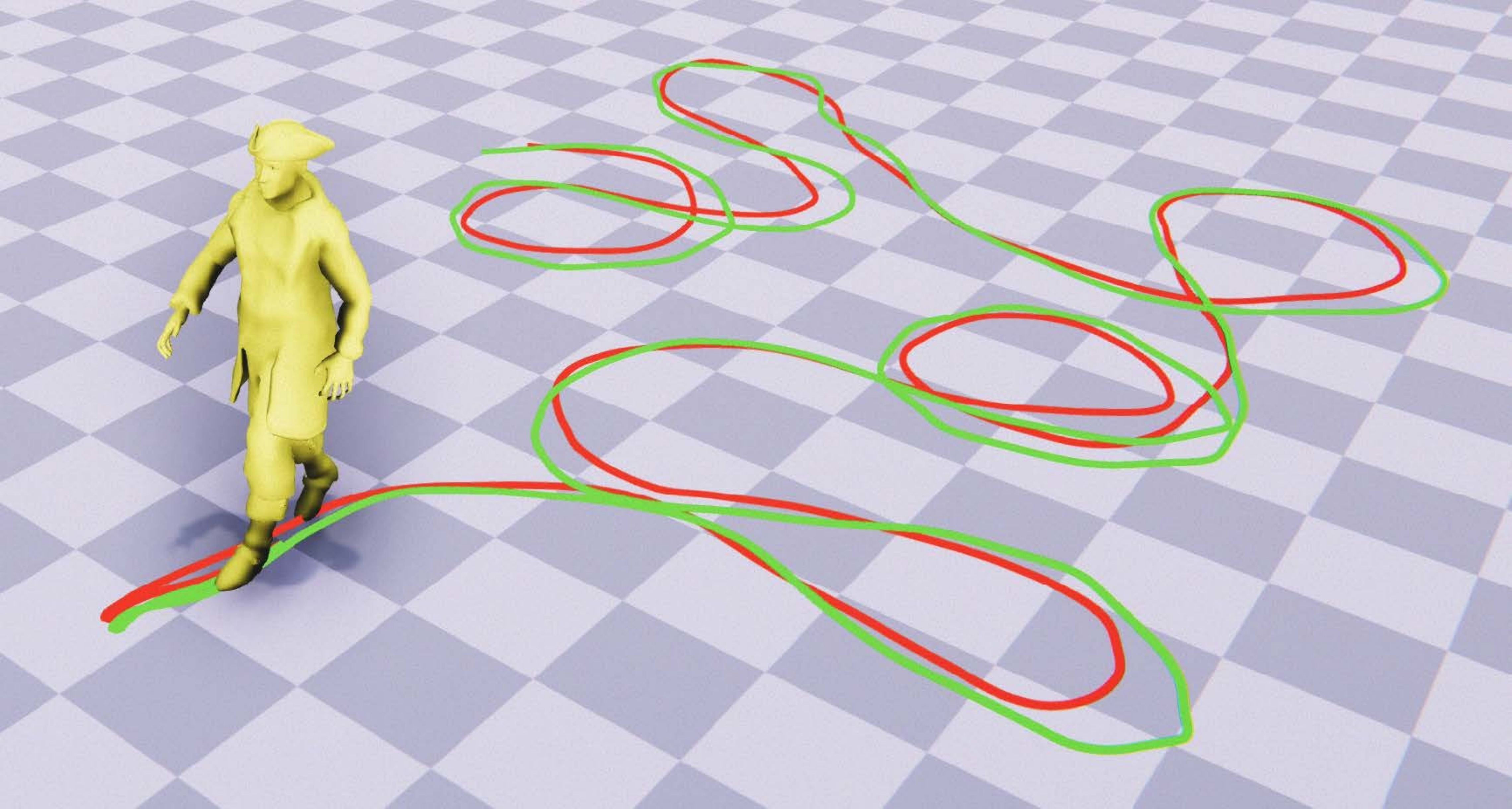}
       \label{fig:pfnn_traj}
    }
    \subfloat[TPTN]{
        \includegraphics[width=0.47\linewidth]{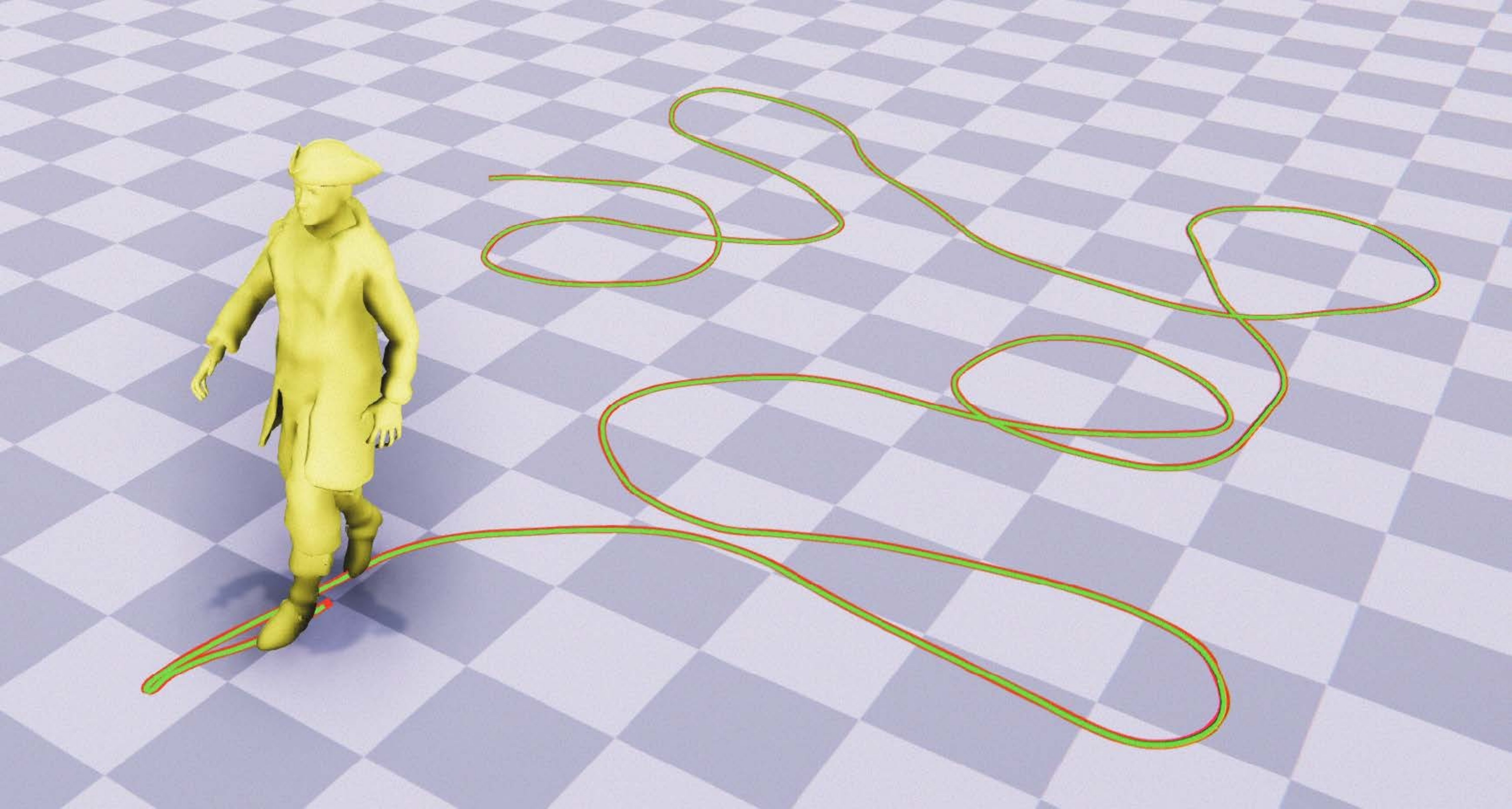}
        \label{fig:tptn_on_pfnn}
    }
  \caption{Trajectory-following results for PFNN and TPTN trained on PFNN's dataset (from 4m38s to 4m58s in the accompanying video). The input and synthesized trajectories are in red and green, respectively. IK is disabled in this experiment.}
  \label{fig:pfnn_vs_tptn}
\end{figure}

\begin{table}[t]
  \centering
  \resizebox{0.33\textwidth}{!}{
    \begin{tabular}{cccc}
\toprule[0.7pt]
               & GT     & PFNN           & TPTN            \\ \midrule
BM (deg/frame) & 70.060 & 82.136         & \textbf{67.473} \\ \midrule
AFS (cm/frame) & 0.184  & {0.354} & \textbf{0.215}           \\ \midrule
SSIM           & 1.0    & 0.977          & \textbf{0.986} \\ \midrule
MPJPE           & 0.0    & \textbf{4.484}          & {7.318}  \\ \bottomrule[0.7pt]
\end{tabular}
  }
  \caption{Comparisons on the metrics for PFNN dataset. GT: ground-truth mocap data. IK is disabled in this experiment. Please refer to the supplementary material for these metrics on different types of motions.}
  \label{tab:pfnn_comps}
\end{table}

To compare with PFNN, we exploit its released codes and trained model weights to synthesize motions using the input vectors extracted from the test dataset. Afterward, we compute the metrics and report them in Tab.~\ref{tab:pfnn_comps}. As illustrated in Tab.~\ref{tab:pfnn_comps}, TPTN outperforms PFNN on BM, AFS, and SSIM with a larger margin, demonstrating that TPTN can generalize to other datasets well.

We also use the two models to synthesize motions following user-specified trajectories. The trajectory distances of results in Fig.~\ref{fig:pfnn_vs_tptn} are 8.82cm/frame with PFNN and 0.441cm/frame with TPTN, respectively, which demonstrates that TPTN can generate motion following the input trajectory more accurately.

\begin{table}[t]%
  \centering
    \resizebox{0.48\textwidth}{!}{
      \begin{tabular}{cccc}
        \toprule[1pt]
        VS                                                                     & \begin{tabular}[c]{@{}c@{}}MVAE\\ (mean: 2.390\\ std: 0.507)\end{tabular}     & \begin{tabular}[c]{@{}c@{}}HTSS\\ (mean: 2.448\\ std: 0.403)\end{tabular}         & \begin{tabular}[c]{@{}c@{}}NeuralLayering\\ (mean: 3.545\\ std: 0.532)\end{tabular} \\ \midrule
        \begin{tabular}[c]{@{}c@{}}TPTN\\ (mean: 3.922\\ std: 0.416)\end{tabular} & \begin{tabular}[c]{@{}c@{}}P-value: 9.539e-8\\ t-value: 7.747\end{tabular} & \begin{tabular}[c]{@{}c@{}}P-value: 2.508e-8\\ t-value: 8.443\end{tabular} & \begin{tabular}[c]{@{}c@{}}P-value: 3.757e-2\\ t-value: 1.850\end{tabular}       \\ \bottomrule[1pt]
      \end{tabular}
    }
    \caption{T-test of user-study results for comparisons on non-dance motions (confidence interval=0.95). MVAE, HTSS, and NeuralLayering are trained on the dataset without dances. VS: performing t-test between the results of TPTN and all the results of baseline models in the second row. Mean: the average scores of generated sequences rated by all the participants compared to mocap sequences in the same group. Std: the standard deviation of the rated scores.}
    \label{tab:comparisons-t-test}
\end{table}

\begin{table}[t]%
  \centering
    \resizebox{0.38\textwidth}{!}{
      \begin{tabular}{ccc}
        \toprule[0.8pt]
        VS                                                                     & \begin{tabular}[c]{@{}c@{}}MVAE\\ (mean: 2.709\\ std: 0.592)\end{tabular}     & \begin{tabular}[c]{@{}c@{}}HTSS\\ (mean: 2.549\\ std: 0.427)\end{tabular}        \\ \midrule
        \begin{tabular}[c]{@{}c@{}}TPTN\\ (mean: 3.907\\ std: 0.407)\end{tabular} & \begin{tabular}[c]{@{}c@{}}P-value: 1.634e-6\\ t-value: 6.017\end{tabular} & \begin{tabular}[c]{@{}c@{}}P-value: 8.180e-9\\ t-value: 8.300\end{tabular}       \\ \bottomrule[0.8pt]
      \end{tabular}
    }
    \caption{T-test of user-study results for comparisons on all test motions containing dances (confidence interval=0.95). MVAE and HTSS are trained on the same dataset as TPTN. VS: performing t-test between the results of TPTN and all the results of baseline models in the second row. Mean: the average scores of generated sequences rated by all the participants compared to mocap sequences in the same group. Std: the standard deviation of the rated scores.}%
    \label{tab:comp-dance-t-test}%
\end{table}

\subsection{User study}
\label{sec:user_study}
To evaluate the visual quality of the generated motions of TPTN, MVAE, HTSS, and NeuralLayering, we conducted a user study to obtain subjective judgments. We first synthesize 11 motion sequences for TPTN and baseline models trained on the dataset without dances as described in Section.~\ref{sec:comparison}, but at most 600 frames for each sequence. Consequently, the average length of the generated sequences is about 10 seconds.

First, we present 14 participants with 11 groups of motion videos. Each group contains five motion sequences: the ground-truth mocap data as reference, and the rest four sequences are generated by TPTN as well as the baseline models. Second, we ask them to rank four videos rendered with the generated motions with respect to the reference motion from 5 aspects. We quantify the user study as a preference score. It rates the overall quality of the generated motions from 1 (least low quality) to 5 (most high quality). The participants include four females and ten males. All of them have experience with 3D animation or games. More details of the user study are reported in the supplementary material.

We perform a t-test on the user study results to verify the hypothesis that the TPTN can generate motions of better quality than baseline models. The results are shown in Tab.~\ref{tab:comparisons-t-test}. The P-values of TPTN vs. other baseline models are all less than the selected threshold (0.05). Therefore, the motions generated by TPTN are significantly different from those generated by baseline models. The average scores of motion sequences rated by the participants (the mean values in Tab.~\ref{tab:comparisons-t-test}) for TPTN are higher than other baseline models. It verifies that the TPTN can generate better motions than baseline models in this user study. Fig.~\ref{fig:slidings}\&~\ref{fig:user_study} show examples of generated motion frames used in this study. In Fig.~\ref{fig:user_study}, when the character twists, the arms and feet in the MVAE-generated motion and the feet in the HTSS-generated motion are in the wrong positions. The right arm in the NeuralLayering-generated motion bends a bit more unnaturally than that in the TPTN-generated motion.

We also conduct a similar user study for TPTN, MVAE and HTSS trained on the same dataset as TPTN. The videos presented to the participants contain 13 groups with two test dance sequences. The t-test results in Tab.~\ref{tab:comp-dance-t-test} verify that the TPTN can generate motions, even for dances, of better quality than baseline models.

\subsection{Evaluations on different experiment settings}
\label{sec:evaluations}
Considering the large number of evaluation experiments and the limited computation resources available to us, it is too expensive to evaluate various hyper-parameters by re-training the model on the full training dataset. Without loss of generality, we only use a subset and call it $\boldsymbol{D_{L}}$, to conduct all the evaluation experiments. The training subset of $\boldsymbol{D_{L}}$ consists of 39 sequences randomly selected from the full dataset, it contains all the first 15 types of motions and transitions between any two kinds of motions, and the test subset of it comprises nine sequences containing all the first 15 types of motions. As a result, the training and test subsets consist of 40,712 and 8,641 frames. If not mentioned, all models are trained and tested on $\boldsymbol{D_{L}}$ hereafter.

\begin{table}[]
    \centering
    \resizebox{0.48\textwidth}{!}{
        \begin{tabular}{ccccccc}
\toprule[1pt]
                & GT     & EBTN   & FPTN-R         & FPTN-S & 5PTN   & TPTN            \\ \midrule
Number of parts & 1      & 1      & 4              & 4      & 5      & 2               \\ \midrule
Overlap joints   & -      & -      & root           & spines  & root   & root            \\ \midrule
BM (deg/frame)  & 88.975 & 79.207 & 78.350         & 80.408 & 79.672 & \textbf{82.162} \\ \midrule
AFS (cm/frame)  & 0.269  & 0.348  & \textbf{0.323} & 0.342  & 0.363  & 0.340           \\ \midrule
SSIM            & 1.0    & 0.991  & 0.990          & 0.989  & 0.991  & \textbf{0.991}  \\ \bottomrule[1pt]
\end{tabular}
    }
    \caption{The three metrics on the test subset for body parts evaluation (the full table can be found in the supplementary material). GT: ground-truth mocap data. IK is disabled in this experiment. BM and SSIM of motions generated by TPTN are better than the other four models.}
    \label{tab:evaluation_body_parts}
\end{table}

\noindent\textbf{Why two body parts}. To verify the effectiveness of the two-part based design, we conduct experiments as follows:
\begin{itemize}
    \item We remove the two-part separation and use the same ARMs to train a network on the motion data of the entire human body and call this model EBTN.
    \item We partition the human body into four parts, i.e., two arms and two legs, and all four parts contain the spine. We model the four parts using four streams constructed by the same ARMs as TPTN and train two different models: 1) the overlapped data is the root-joint information same as TPTN, and we call this model FPTN-R; 2) the overlapped data is the data representation vectors of all the joints in the spine, and we call this model FPTN-S.
    \item We partition the human body into five parts, i.e., two arms, two legs, and the spine. Similarly, we use five streams to model these parts and call this model 5PTN.
\end{itemize}
The parameters of EBTN, FPTN-R, FPTN-S, and 5PTN are \textasciitilde 3.1M, \textasciitilde 3.37M, \textasciitilde 3.38M and \textasciitilde 3.4M respectively. For each body part in FPTN-R, FPTN-S, and 5PTN, we fuse all the other parts' features using a similar design as TPTN's FFL. We can see, from Tab.~\ref{tab:evaluation_body_parts}, that TPTN outperforms other models on the metrics, which demonstrates the superiority of the TPTN. Moreover, their inference speed is far slower(\textasciitilde 37fps for FPTN-R, \textasciitilde 37fps for FPTN-S, and \textasciitilde 29fps for 5PTN). We hypothesize that decomposing the human body into four or five parts makes it harder for our lightweight FFL and consistency loss to capture the correlations among parts and lose some global information about the entire body.

Although EBTN achieves decent results for some types of motions, it fails to model motions like twists. With EBTN, the character performs a wrong type of pose when given a ``twist'' label (refer to the supplementary material for the result and the accompanying video from 5m04s to 5m14s for the motion). Therefore, it verifies the importance of decomposing the human body into two parts for modeling a broader scope of complex heterogeneous motions. For FPTN-R, FPTN-S, and 5PTN, there are no wrong poses like EBTN, which implies that decomposing the skeleton into parts in motion modeling is beneficial to reducing the complexity and difficulty of modeling heterogeneous motions.

\begin{table}[]
    \centering
    \resizebox{0.46\textwidth}{!}{
        \begin{tabular}{cccccc}
\toprule[1pt]
               & GT     & w/o-FFL & CrossAttn & w/o-CL         & TPTN            \\ \midrule
BM (deg/frame) & 88.975 & 79.107  & 76.955    & 78.124         & \textbf{82.162} \\ \midrule
AFS (cm/frame) & 0.269  & 0.345   & 0.384     & \textbf{0.330} & 0.340           \\ \midrule
SSIM           & 1.0    & 0.991   & 0.990     & 0.990          & \textbf{0.991}  \\ \bottomrule[1pt]
\end{tabular}
    }
    \caption{The three metrics on the test subset for feature fusion and consistency loss evaluation (the full table can be found in the supplementary material). GT: ground-truth mocap data. IK is disabled in this experiment. BM and SSIM of motions generated by TPTN are better than the other three models.}
    \label{tab:ffl_ablations}
\end{table}

\noindent\textbf{Feature fusion layer and consistency loss}. In this experiment, we first remove the FFL from TPTN and call this model w/o-FFL. As shown in Tab.~\ref{tab:ffl_ablations}, all metrics worsen. The arms and legs may move on the same side in the generated motions of this model (refer to the supplementary material and the accompanying video 5m54s to 6m05s for the results). We also replace the FFL with the cross-attention layer as in~\cite{vaswani2017attention} to determine the influence of the FFL's structure and name this model as CrossAttn. When fusing the upper body part with the lower body part, the features of the lower body part are input to the cross-attention layer as the query vectors, and the features of the upper body part as the key and value vectors. Instead, when fusing the lower body part with the upper body part, the query vectors are the features of the upper body part, and the key and value vectors are the features of the lower body parts. The metrics of motions generated by this trained model downgrade as illustrated in Tab.~\ref{tab:ffl_ablations}. Since our implementation of the cross-attention mechanism mainly computes attention weights of features at different frames rather than fuses the features of these two body parts, it might be the reason that attention only can not serve the purpose of feature fusion in our model well.

We then discard the consistency loss (CL) from the training losses and denote this model by w/o-CL. The metrics in Tab.~\ref{tab:ffl_ablations} indicate a decline in performance. To figure out why the BM values decrease with a large margin compared to TPTN, we computed the body movement of upper and lower body parts in the test dataset using ground-truth data, TPTN-generated data, and w/o-CL-generated data. The BM values of the upper body joints for ground-truth, TPTN-generated and w/o-CL-generated motions are 64.692, 59.084, and 55.274, respectively. Similarly, the BM values of lower body motions are 43.645, 38.165, and 37.234, respectively. We observe a more significant decrease in BM values in the upper body motions compared to the lower body motions, which is the primary reason for the overall BM decrease in this model. We hypothesize that the lower body information can aid the prediction of the upper body motion more effectively than the upper body information aids the lower body motion. Therefore, without CL, it's challenging for the network to capture the details of the upper body motions. In contrast, without the upper body information passed in CL, the network can still generate lower body motions of comparable quality, as the BM value of the lower body part does not drop too much. In addition, the accompanying video from 5m54s to 6m05s illustrates that the character's left arm swings more naturally when CL is present during the transition from running back to walking. 

The AFS metric is computed based on the motions of the feet and toes, which are primarily influenced by the lower body's motion. Hence, if the generated lower body motions are accurate, the AFS value will be low. We observe that the ARMs for the lower body part work well to generate realistic lower body motions. As a result, our experimental results show that the lower body motions generated without FFL and CL modules are accurate enough to achieve comparable AFS performance to TPTN. However, these two modules enhance the visual quality of the generated motions, as illustrated in Fig. 3 in the supplementary material and the accompanying video from 5m54s to 6m05s. Additionally, as demonstrated in the previous two paragraphs, FFL and CL significantly improve the BM metric. Therefore, we can conclude that FFL and CL are necessary for TPTN.

\begin{table}[]
    \centering
    \resizebox{0.48\textwidth}{!}{
        \begin{tabular}{ccccccc}
\toprule[1pt]
               & GT     & TRL39  & TRL135 & TP-NLNN & TP-LSTM4LR & TPTN            \\ \midrule
BM (deg/frame) & 88.975 & 81.952 & 77.476 & 81.467  & 106.122    & \textbf{82.162} \\ \midrule
AFS (cm/frame) & 0.269  & 0.361  & 0.345  & 0.732   & 0.541      & \textbf{0.340}  \\ \midrule
SSIM           & 1.0    & 0.990  & 0.991  & 0.983   & 0.979      & \textbf{0.991}  \\ \bottomrule[1pt]
\end{tabular}
    }
    \caption{The three metrics on the test subset for temporal receptive field length and auto-regressive module backbone evaluation (the full table can be found in the supplementary material). GT: ground-truth mocap data. IK is disabled in this experiment. The metrics of motions generated by TPTN are the best.}
    \label{tab:trl_backbone_ablations}
\end{table}

\noindent\textbf{Temporal receptive field (TRL) length}. Choosing a suitable $TRL$ for the auto-regressive network is critical to synthesizing realistic and diverse motions. To evaluate our TRL selection, namely 87, we set the $TRL$ to be 39 and 135 by setting the temporal window length $lm$ to be 4 and 12. Then, we evaluate the two models (TRL39 and TRL135) on the metrics after training them from scratch. Finally, all scores are worse than that obtained when $TRL=87$, as illustrated in Tab.~\ref{tab:trl_backbone_ablations}. If the $TRL$ is too short, the network can only observe limited context information and may generate unnatural poses. While if the $TRL$ is too long, the generated motions may be over-smoothed and result in relatively small body movement. Please refer to the accompanying video from 6m35s to 6m46s for the results.

\noindent\textbf{Why Transformer}. To evaluate the transformer's superiority in modeling heterogeneous human motions, we replaced the multi-head attention layers in each ARM with non-local neural networks (NLNN)~\cite{wang2018non} and four-layer LSTMs (LSTM4LR) similar to the student policy network in~\cite{lee2021learning}. We selected ResNet50 to construct the NLNN based on the experimental results of ResNets with different layers. We named the resulting models TP-NLNN and TP-LSTM4LR, respectively. Please refer to the supplementary material for the detailed network architectures. We trained TP-NLNN and TP-LSTM4LR using the same settings as TPTN and evaluated their performance using the same test dataset and metrics. The result values for TP-NLNN and TP-LSTM4LR are reported in Tab.~\ref{tab:trl_backbone_ablations}, and our TPTN performs the best in all the scores.

We think the main reason that the transformer outperforms convolutional neural networks (CNN) and recurrent neural networks (RNN) (including LSTM) is that the transformer can better capture the temporal dependencies among human motions. It's known that RNNs suffer from long dependency issues, which are inherently related to their recursion attribute. Although LSTM and bi-directional RNN models can mitigate this problem in many time series-related tasks, including locomotion modeling, we still encounter some cases where the generated motions tend to converge to mean poses or fail to perform complex poses, like whirlwind kicking, especially in our heterogeneous human motion synthesis task. As for CNN, it employs kernels of different sizes to capture local dependencies among frames within various window sizes. Stacks of CNN layers are used to expand the temporal receptive field length, such as~\cite{kim-2014-convolutional, hou2021causal}. However, determining appropriate window sizes to better capture local dependencies, particularly for different types of complex human motions, is tricky. Consequently, selecting optimal kernel sizes for CNN in such tasks is a significant challenge. In contrast, the transformer processes the sequence as a whole through multi-head attention layers and positional encodings, avoiding the recursion in RNN and the determination of kernel sizes that correspond to the local temporal dependencies in CNN. As a result, the transformer can adaptively learn such dependencies from the datasets during training. The statistics in Tab.~\ref{tab:trl_backbone_ablations} verify the effectiveness of the transformer in modeling heterogeneous human motions in these controllable motion synthesis experiments. In addition, the trained TP-NLNN and TP-LSTM4LR network may fail to generate complex motions like whirlwind kicking and produce more jitters in the generated motions (refer to the accompanying video from 6m51s to 7m44s for the video comparisons).

\begin{table}[]
    \centering
    \resizebox{0.48\textwidth}{!}{
        \begin{tabular}{ccccccccc}
            \toprule[1pt]
            Metrics        & GT     & w/o-FK & w/o-OS & w/o-C  & L2 Loss & w/o-F & TPTN            \\ \midrule
            BM (deg/frame) & 88.975 & 79.075 & 80.826 & 80.426 & 55.328 & \textbf{89.086}  & { 82.162} \\ \midrule
            AFS (cm/frame) & 0.269  & { 0.342}  & 0.343  & 0.348  & 0.434  & 0.551 & \textbf{0.340}  \\ \midrule
            SSIM           & 1.0      & 0.991  & 0.990  & 0.990  & 0.991 & 0.989  & \textbf{0.991}  \\ \bottomrule[1pt]
        \end{tabular}
    }
    \caption{Ablation study results of the three metrics on the test subset (the full table can be found in the supplementary material). GT: ground-truth mocap data. IK is disabled in this experiment. AFS and SSIM of motions generated by TPTN are better than the other five models.}
    \label{tab:other_ablations}
\end{table}

\noindent\textbf{Other ablation studies}. We first conducted four additional experiments to further investigate the impact of various components on the performance of our model. The experimental settings are: 1) discarding the FK layer and FK loss (w/o-FK); 2) removing the oversampling technique (w/o-OS), 3) replacing the 1D causal convolutions before the ARMs (w/o-C) with 1D convolutions (kernel size being 1); 4) replacing the Gaussian loss with L2 loss (L2 Loss);  Overall, TPTN outperforms its variants on the metrics, which means that our TPTN can synthesize more agile and smoother motions than its variants. It's worth noting that the L2 loss assigns the same weight to different elements in the motion representation vector $\mathcal{X}_{n}$, which fails to differentiate the importance of different elements. In contrast, the covariance matrix in the Gaussian loss effectively encodes the importance of each element. Therefore, replacing Gaussian loss with L2 loss over-smoothes the generated motions. Please refer to the accompanying video from 7m49s to 8m11s for these results.

Additionally, we remove the foot contact loss (w/o-F) as well as the second term in FK loss to investigate its influence on the generated motion. We have to remove the second term in FK loss in this experiment since it relies on the foot contact label predicted by the foot contact loss. The synthesized motion of the w/o-F variant has severe foot sliding artifacts, resulting in a significantly worse AFS value of 0.551 compared to 0.340, the AFS value of TPTN, as shown in Tab.~\ref{tab:other_ablations}. This result verifies that the foot contact loss is necessary to eliminate the foot sliding, which is consistent with the ablation study result of the foot contact loss in ~\cite{shi2020motionet}. Note that the BM increase with the w/o-F variant is primarily due to the jitters of the generated motions. Please refer to the accompanying video from 8m17s to 8m39s for the video comparisons.

\noindent\textbf{Where and how to input the control signals}. Our first try was to concatenate features of control signals into the input of the first multi-head attention layer in each auto-regressive module. However, we found that the predicted control signals $\mathbf{c}_n^p$ and $\mathbf{c}_n^d$ quickly exploded in a few frames. We calculated the derivatives of the network's output with input control signals to find the reason. We found that the derivatives were always around 150 and far bigger than the derivatives of output over input motion features, which were all around 1. We then removed  motion types control signal $\mathbf{c}_n^t$ from the concatenation each time and trained the model from scratch. We noticed that the derivatives of control signals drastically decreased from around 150 to around 10. We hypothesize that it is because we do not normalize one-hot motion-type vectors since their values are already in the range [0,1], which is different from other normalized control signals. We thus weaken the impact of control signals on the network by removing them from the first multi-head attention layer. Finally, as shown in Fig.~\ref{fig:ars}, we concatenate features of control signals with the motion features before the multi-head attention layer. This way, the derivatives decrease to around 4, making our network able to synthesize natural motions from its own predicted control signals $\mathbf{c}_n^p$ and $\mathbf{c}_n^d$.

\section{Conclusion}
We have designed a novel two-part auto-regressive model, i.e., TPTN, by exploiting attention mechanisms to synthesize high-quality motions of different types like heterogeneous gymnastic motions, locomotion, and dances. With the help of the introduced consistency loss and feature fusion layer, the TPTN can make the upper and lower body parts coordinate well with each other in the generated motions. Heterogeneous motions, like cartwheels, can be synthesized in real-time after training the system with mocap data of these types.

\noindent\textbf{Limitation}.
As stated in Section~\ref{sec:motion_results}, it is not easy to accurately control the trajectories of motions in the air. Our TPTN cannot control the spinning speed of the character in the air and how long the character stays in the air. Therefore, we plan to enrich the variations of such motions like side somersault and whirlwind kicking by capturing more motions of these types. We also plan to investigate data augmentation to enhance the variations of such motions. Another problem is that the TPTN trained on our dataset performs poorly in motions comprising massive interactions with environments like sitting or lying. For instance, if the character frequently lies or rolls on the ground during dance performances, our TPTN will generate motions with apparent body-ground penetration (see the video named tptn\_supplementary.mp4 for details). The penetration might be mitigated by adding a loss to explicitly constrain all joints to be above the support plane.

\noindent\textbf{Future work}. It is interesting to investigate how to realize fine-grained control on key joints like~\cite{starke2021neural}. Simply assigning a general type label like ``ballet'' will bring ambiguities to the dance poses. Therefore, we plan to consult with professional artists to give more exact type labels to dance rhythm or beats. We are also interested in controlling dance generation by user input control signals and music simultaneously. In addition, we plan to capture more dances of different characters and exploit the oversampling technique on the skeleton sizes to improve the quality of generated dances.


%
%
%
%
%
%
%
%

\ifCLASSOPTIONcaptionsoff
  \newpage
\fi

%
%
%
%

\bibliographystyle{IEEEtran}
\bibliography{tptn_jrnl}

\begin{thebibliography}{10}
\providecommand{\url}[1]{#1}
\csname url@samestyle\endcsname
\providecommand{\newblock}{\relax}
\providecommand{\bibinfo}[2]{#2}
\providecommand{\BIBentrySTDinterwordspacing}{\spaceskip=0pt\relax}
\providecommand{\BIBentryALTinterwordstretchfactor}{4}
\providecommand{\BIBentryALTinterwordspacing}{\spaceskip=\fontdimen2\font plus
\BIBentryALTinterwordstretchfactor\fontdimen3\font minus
  \fontdimen4\font\relax}
\providecommand{\BIBforeignlanguage}[2]{{%
\expandafter\ifx\csname l@#1\endcsname\relax
\typeout{** WARNING: IEEEtran.bst: No hyphenation pattern has been}%
\typeout{** loaded for the language `#1'. Using the pattern for}%
\typeout{** the default language instead.}%
\else
\language=\csname l@#1\endcsname
\fi
#2}}
\providecommand{\BIBdecl}{\relax}
\BIBdecl

\bibitem{holden2017phase}
D.~Holden, T.~Komura, and J.~Saito, ``Phase-functioned neural networks for
  character control,'' \emph{ACM Transactions on Graphics (TOG)}, vol.~36,
  no.~4, pp. 1--13, 2017.

\bibitem{ling2020character}
H.~Y. Ling, F.~Zinno, G.~Cheng, and M.~Van De~Panne, ``Character controllers
  using motion vaes,'' \emph{ACM Transactions on Graphics (TOG)}, vol.~39,
  no.~4, pp. 40--1, 2020.

\bibitem{lee2018interactive}
K.~Lee, S.~Lee, and J.~Lee, ``Interactive character animation by learning
  multi-objective control,'' \emph{ACM Transactions on Graphics (TOG)},
  vol.~37, no.~6, pp. 1--10, 2018.

\bibitem{starke2020local}
S.~Starke, Y.~Zhao, T.~Komura, and K.~Zaman, ``Local motion phases for learning
  multi-contact character movements,'' \emph{ACM Transactions on Graphics
  (TOG)}, vol.~39, no.~4, pp. 54--1, 2020.

\bibitem{lee2021learning}
K.~Lee, S.~Min, S.~Lee, and J.~Lee, ``Learning time-critical responses for
  interactive character control,'' \emph{ACM Transactions on Graphics (TOG)},
  vol.~40, no.~4, pp. 1--11, 2021.

\bibitem{lee2021learning2}
S.~Lee, S.~Lee, Y.~Lee, and J.~Lee, ``Learning a family of motor skills from a
  single motion clip,'' \emph{ACM Transactions on Graphics (TOG)}, vol.~40,
  no.~4, pp. 1--13, 2021.

\bibitem{starke2021neural}
S.~Starke, Y.~Zhao, F.~Zinno, and T.~Komura, ``Neural animation layering for
  synthesizing martial arts movements,'' \emph{ACM Transactions on Graphics
  (TOG)}, vol.~40, no.~4, pp. 1--16, 2021.

\bibitem{chen2020dynamic}
W.~Chen, H.~Wang, Y.~Yuan, T.~Shao, and K.~Zhou, ``Dynamic future net:
  Diversified human motion generation,'' in \emph{Proceedings of the 28th ACM
  International Conference on Multimedia}, 2020, pp. 2131--2139.

\bibitem{perez2021transflower}
G.~V. Perez, J.~Beskow, G.~Henter, A.~Holzapfel, P.-Y. Oudeyer, and
  S.~Alexanderson, ``Transflower: probabilistic autoregressive dance generation
  with multimodal attention,'' in \emph{SIGGRAPH Asia 2021-14th ACM SIGGRAPH
  Conference and Exhibition on Computer Graphics and Interactive Techniques},
  2021.

\bibitem{fragkiadaki2015recurrent}
K.~Fragkiadaki, S.~Levine, P.~Felsen, and J.~Malik, ``Recurrent network models
  for human dynamics,'' in \emph{Proceedings of the IEEE International
  Conference on Computer Vision}, 2015, pp. 4346--4354.

\bibitem{wang2019combining}
Z.~Wang, J.~Chai, and S.~Xia, ``Combining recurrent neural networks and
  adversarial training for human motion synthesis and control,'' \emph{IEEE
  transactions on visualization and computer graphics}, vol.~27, no.~1, pp.
  14--28, 2019.

\bibitem{henter2020moglow}
G.~E. Henter, S.~Alexanderson, and J.~Beskow, ``Moglow: Probabilistic and
  controllable motion synthesis using normalising flows,'' \emph{ACM
  Transactions on Graphics (TOG)}, vol.~39, no.~6, pp. 1--14, 2020.

\bibitem{starke2019neural}
S.~Starke, H.~Zhang, T.~Komura, and J.~Saito, ``Neural state machine for
  character-scene interactions.'' \emph{ACM Trans. Graph.}, vol.~38, no.~6, pp.
  209--1, 2019.

\bibitem{wang2019spatio}
H.~Wang, E.~S. Ho, H.~P. Shum, and Z.~Zhu, ``Spatio-temporal manifold learning
  for human motions via long-horizon modeling,'' \emph{IEEE transactions on
  visualization and computer graphics}, vol.~27, no.~1, pp. 216--227, 2019.

\bibitem{liu2021aggregated}
Z.~Liu, K.~Lyu, S.~Wu, H.~Chen, Y.~Hao, and S.~Ji, ``Aggregated multi-gans for
  controlled 3d human motion prediction,'' in \emph{Proceedings of the AAAI
  Conference on Artificial Intelligence}, vol.~35, no.~3, 2021, pp. 2225--2232.

\bibitem{ghosh2021synthesis}
A.~Ghosh, N.~Cheema, C.~Oguz, C.~Theobalt, and P.~Slusallek, ``Synthesis of
  compositional animations from textual descriptions,'' \emph{arXiv preprint
  arXiv:2103.14675}, 2021.

\bibitem{vaswani2017attention}
A.~Vaswani, N.~Shazeer, N.~Parmar, J.~Uszkoreit, L.~Jones, A.~N. Gomez,
  {\L}.~Kaiser, and I.~Polosukhin, ``Attention is all you need,'' in
  \emph{Advances in neural information processing systems}, 2017, pp.
  5998--6008.

\bibitem{huang2020dance}
R.~Huang, H.~Hu, W.~Wu, K.~Sawada, M.~Zhang, and D.~Jiang, ``Dance revolution:
  Long-term dance generation with music via curriculum learning,'' \emph{arXiv
  preprint arXiv:2006.06119}, 2020.

\bibitem{xia2015realtime}
S.~Xia, C.~Wang, J.~Chai, and J.~Hodgins, ``Realtime style transfer for
  unlabeled heterogeneous human motion,'' \emph{ACM Transactions on Graphics
  (TOG)}, vol.~34, no.~4, pp. 1--10, 2015.

\bibitem{won2021control}
J.~Won, D.~Gopinath, and J.~Hodgins, ``Control strategies for physically
  simulated characters performing two-player competitive sports,'' \emph{ACM
  Transactions on Graphics (TOG)}, vol.~40, no.~4, pp. 1--11, 2021.

\bibitem{aberman2020skeleton}
K.~Aberman, P.~Li, D.~Lischinski, O.~Sorkine-Hornung, D.~Cohen-Or, and B.~Chen,
  ``Skeleton-aware networks for deep motion retargeting,'' \emph{ACM
  Transactions on Graphics (TOG)}, vol.~39, no.~4, pp. 62--1, 2020.

\bibitem{liu2018learning}
L.~Liu and J.~Hodgins, ``Learning basketball dribbling skills using trajectory
  optimization and deep reinforcement learning,'' \emph{ACM Transactions on
  Graphics (TOG)}, vol.~37, no.~4, pp. 1--14, 2018.

\bibitem{wang20143d}
X.~Wang, Q.~Chen, and W.~Wang, ``3d human motion editing and synthesis: A
  survey,'' \emph{Computational and Mathematical methods in medicine}, vol.
  2014, 2014.

\bibitem{jain2016structural}
A.~Jain, A.~R. Zamir, S.~Savarese, and A.~Saxena, ``Structural-rnn: Deep
  learning on spatio-temporal graphs,'' in \emph{Proceedings of the ieee
  conference on computer vision and pattern recognition}, 2016, pp. 5308--5317.

\bibitem{Corona_2020_CVPR}
E.~Corona, A.~Pumarola, G.~Alenya, and F.~Moreno-Noguer, ``Context-aware human
  motion prediction,'' in \emph{IEEE/CVF Conference on Computer Vision and
  Pattern Recognition (CVPR)}, June 2020.

\bibitem{martinez2017human}
J.~Martinez, M.~J. Black, and J.~Romero, ``On human motion prediction using
  recurrent neural networks,'' in \emph{Proceedings of the IEEE Conference on
  Computer Vision and Pattern Recognition}, 2017, pp. 2891--2900.

\bibitem{li2017auto}
Z.~Li, Y.~Zhou, S.~Xiao, C.~He, Z.~Huang, and H.~Li, ``Auto-conditioned
  recurrent networks for extended complex human motion synthesis,'' \emph{arXiv
  preprint arXiv:1707.05363}, 2017.

\bibitem{butepage2017deep}
J.~Butepage, M.~J. Black, D.~Kragic, and H.~Kjellstrom, ``Deep representation
  learning for human motion prediction and classification,'' in
  \emph{Proceedings of the IEEE conference on computer vision and pattern
  recognition}, 2017, pp. 6158--6166.

\bibitem{holden2016deep}
D.~Holden, J.~Saito, and T.~Komura, ``A deep learning framework for character
  motion synthesis and editing,'' \emph{ACM Transactions on Graphics (TOG)},
  vol.~35, no.~4, pp. 1--11, 2016.

\bibitem{Li_2020_CVPR}
M.~Li, S.~Chen, Y.~Zhao, Y.~Zhang, Y.~Wang, and Q.~Tian, ``Dynamic multiscale
  graph neural networks for 3d skeleton based human motion prediction,'' in
  \emph{IEEE/CVF Conference on Computer Vision and Pattern Recognition (CVPR)},
  June 2020.

\bibitem{Cui_2020_CVPR}
Q.~Cui, H.~Sun, and F.~Yang, ``Learning dynamic relationships for 3d human
  motion prediction,'' in \emph{IEEE/CVF Conference on Computer Vision and
  Pattern Recognition (CVPR)}, June 2020.

\bibitem{yu2020structure}
P.~Yu, Y.~Zhao, C.~Li, J.~Yuan, and C.~Chen, ``Structure-aware human-action
  generation,'' in \emph{European Conference on Computer Vision}.\hskip 1em
  plus 0.5em minus 0.4em\relax Springer, 2020, pp. 18--34.

\bibitem{jang2022motion}
D.-K. Jang, S.~Park, and S.-H. Lee, ``Motion puzzle: Arbitrary motion style
  transfer by body part,'' \emph{ACM Transactions on Graphics (TOG)}, 2022.

\bibitem{villegas2018neural}
R.~Villegas, J.~Yang, D.~Ceylan, and H.~Lee, ``Neural kinematic networks for
  unsupervised motion retargetting,'' in \emph{Proceedings of the IEEE
  Conference on Computer Vision and Pattern Recognition}, 2018, pp. 8639--8648.

\bibitem{degardin2022generative}
B.~Degardin, J.~Neves, V.~Lopes, J.~Brito, E.~Yaghoubi, and H.~Proen{\c{c}}a,
  ``Generative adversarial graph convolutional networks for human action
  synthesis,'' in \emph{Proceedings of the IEEE/CVF Winter Conference on
  Applications of Computer Vision}, 2022, pp. 1150--1159.

\bibitem{li2022ganimator}
P.~Li, K.~Aberman, Z.~Zhang, R.~Hanocka, and O.~Sorkine-Hornung, ``Ganimator:
  Neural motion synthesis from a single sequence,'' \emph{arXiv preprint
  arXiv:2205.02625}, 2022.

\bibitem{bengio2015scheduled}
S.~Bengio, O.~Vinyals, N.~Jaitly, and N.~Shazeer, ``Scheduled sampling for
  sequence prediction with recurrent neural networks,'' \emph{arXiv preprint
  arXiv:1506.03099}, 2015.

\bibitem{starke2022deepphase}
S.~Starke, I.~Mason, and T.~Komura, ``Deepphase: periodic autoencoders for
  learning motion phase manifolds,'' \emph{ACM Transactions on Graphics (TOG)},
  vol.~41, no.~4, pp. 1--13, 2022.

\bibitem{siyao2022bailando}
L.~Siyao, W.~Yu, T.~Gu, C.~Lin, Q.~Wang, C.~Qian, C.~C. Loy, and Z.~Liu,
  ``Bailando: 3d dance generation via actor-critic gpt with choreographic
  memory,'' in \emph{CVPR}, 2022.

\bibitem{tang2022real}
X.~Tang, H.~Wang, B.~Hu, X.~Gong, R.~Yi, Q.~Kou, and X.~Jin, ``Real-time
  controllable motion transition for characters,'' \emph{arXiv preprint
  arXiv:2205.02540}, 2022.

\bibitem{brown2020language}
T.~B. Brown, B.~Mann, N.~Ryder, M.~Subbiah, J.~Kaplan, P.~Dhariwal,
  A.~Neelakantan, P.~Shyam, G.~Sastry, A.~Askell \emph{et~al.}, ``Language
  models are few-shot learners,'' \emph{arXiv preprint arXiv:2005.14165}, 2020.

\bibitem{baevski2020wav2vec}
A.~Baevski, H.~Zhou, A.~Mohamed, and M.~Auli, ``wav2vec 2.0: A framework for
  self-supervised learning of speech representations,'' \emph{arXiv preprint
  arXiv:2006.11477}, 2020.

\bibitem{dhariwal2020jukebox}
P.~Dhariwal, H.~Jun, C.~Payne, J.~W. Kim, A.~Radford, and I.~Sutskever,
  ``Jukebox: A generative model for music,'' \emph{arXiv preprint
  arXiv:2005.00341}, 2020.

\bibitem{zhou2021informer}
H.~Zhou, S.~Zhang, J.~Peng, S.~Zhang, J.~Li, H.~Xiong, and W.~Zhang,
  ``Informer: Beyond efficient transformer for long sequence time-series
  forecasting,'' in \emph{Proceedings of AAAI}, 2021.

\bibitem{dosovitskiy2020image}
A.~Dosovitskiy, L.~Beyer, A.~Kolesnikov, D.~Weissenborn, X.~Zhai,
  T.~Unterthiner, M.~Dehghani, M.~Minderer, G.~Heigold, S.~Gelly \emph{et~al.},
  ``An image is worth 16x16 words: Transformers for image recognition at
  scale,'' \emph{arXiv preprint arXiv:2010.11929}, 2020.

\bibitem{liu2021swin}
Z.~Liu, Y.~Lin, Y.~Cao, H.~Hu, Y.~Wei, Z.~Zhang, S.~Lin, and B.~Guo, ``Swin
  transformer: Hierarchical vision transformer using shifted windows,''
  \emph{arXiv preprint arXiv:2103.14030}, 2021.

\bibitem{yan2021videogpt}
W.~Yan, Y.~Zhang, P.~Abbeel, and A.~Srinivas, ``Videogpt: Video generation
  using vq-vae and transformers,'' \emph{arXiv preprint arXiv:2104.10157},
  2021.

\bibitem{mao2020history}
W.~Mao, M.~Liu, and M.~Salzmann, ``History repeats itself: Human motion
  prediction via motion attention,'' in \emph{European Conference on Computer
  Vision}.\hskip 1em plus 0.5em minus 0.4em\relax Springer, 2020, pp. 474--489.

\bibitem{li2021ai}
R.~Li, S.~Yang, D.~A. Ross, and A.~Kanazawa, ``Ai choreographer: Music
  conditioned 3d dance generation with aist++,'' in \emph{Proceedings of the
  IEEE/CVF International Conference on Computer Vision}, 2021, pp.
  13\,401--13\,412.

\bibitem{li2020learning}
J.~Li, Y.~Yin, H.~Chu, Y.~Zhou, T.~Wang, S.~Fidler, and H.~Li, ``Learning to
  generate diverse dance motions with transformer,'' \emph{arXiv preprint
  arXiv:2008.08171}, 2020.

\bibitem{adobe:mixamo}
\BIBentryALTinterwordspacing
Adobe, ``Mixamo,'' 2021, [Online; accessed 19-March-2021]. [Online]. Available:
  \url{https://www.mixamo.com/}
\BIBentrySTDinterwordspacing

\bibitem{Grassia_1998}
F.~S. Grassia, ``Practical parameterization of rotations using the exponential
  map,'' \emph{Journal of Graphics Tools}, vol.~3, no.~3, pp. 29--48, jan 1998.

\bibitem{oord2016wavenet}
A.~v.~d. Oord, S.~Dieleman, H.~Zen, K.~Simonyan, O.~Vinyals, A.~Graves,
  N.~Kalchbrenner, A.~Senior, and K.~Kavukcuoglu, ``Wavenet: A generative model
  for raw audio,'' \emph{arXiv preprint arXiv:1609.03499}, 2016.

\bibitem{JMLR:v15:srivastava14a}
N.~Srivastava, G.~Hinton, A.~Krizhevsky, I.~Sutskever, and R.~Salakhutdinov,
  ``Dropout: A simple way to prevent neural networks from overfitting,''
  \emph{Journal of Machine Learning Research}, vol.~15, no.~56, pp. 1929--1958,
  2014.

\bibitem{tompson2015efficient}
J.~Tompson, R.~Goroshin, A.~Jain, Y.~LeCun, and C.~Bregler, ``Efficient object
  localization using convolutional networks,'' in \emph{Proceedings of the IEEE
  conference on computer vision and pattern recognition}, 2015, pp. 648--656.

\bibitem{loshchilov2017decoupled}
I.~Loshchilov and F.~Hutter, ``Decoupled weight decay regularization,''
  \emph{arXiv preprint arXiv:1711.05101}, 2017.

\bibitem{ling1998data}
C.~X. Ling and C.~Li, ``Data mining for direct marketing: Problems and
  solutions.'' in \emph{Kdd}, vol.~98, 1998, pp. 73--79.

\bibitem{zhang2018mode}
H.~Zhang, S.~Starke, T.~Komura, and J.~Saito, ``Mode-adaptive neural networks
  for quadruped motion control,'' \emph{ACM Transactions on Graphics (TOG)},
  vol.~37, no.~4, pp. 1--11, 2018.

\bibitem{wang2004image}
Z.~Wang, A.~C. Bovik, H.~R. Sheikh, and E.~P. Simoncelli, ``Image quality
  assessment: from error visibility to structural similarity,'' \emph{IEEE
  transactions on image processing}, vol.~13, no.~4, pp. 600--612, 2004.

\bibitem{ionescu2013human3}
C.~Ionescu, D.~Papava, V.~Olaru, and C.~Sminchisescu, ``Human3. 6m: Large scale
  datasets and predictive methods for 3d human sensing in natural
  environments,'' \emph{IEEE transactions on pattern analysis and machine
  intelligence}, vol.~36, no.~7, pp. 1325--1339, 2013.

\bibitem{peng2018deepmimic}
X.~B. Peng, P.~Abbeel, S.~Levine, and M.~van~de Panne, ``Deepmimic:
  Example-guided deep reinforcement learning of physics-based character
  skills,'' \emph{ACM Transactions on Graphics (TOG)}, vol.~37, no.~4, pp.
  1--14, 2018.

\bibitem{holden2020learned}
D.~Holden, O.~Kanoun, M.~Perepichka, and T.~Popa, ``Learned motion matching,''
  \emph{ACM Transactions on Graphics (TOG)}, vol.~39, no.~4, pp. 53--1, 2020.

\bibitem{wang2018non}
X.~Wang, R.~Girshick, A.~Gupta, and K.~He, ``Non-local neural networks,'' in
  \emph{Proceedings of the IEEE conference on computer vision and pattern
  recognition}, 2018, pp. 7794--7803.

\bibitem{kim-2014-convolutional}
\BIBentryALTinterwordspacing
Y.~Kim, ``Convolutional neural networks for sentence classification,'' in
  \emph{Proceedings of the 2014 Conference on Empirical Methods in Natural
  Language Processing ({EMNLP})}.\hskip 1em plus 0.5em minus 0.4em\relax Doha,
  Qatar: Association for Computational Linguistics, Oct. 2014, pp. 1746--1751.
  [Online]. Available: \url{https://aclanthology.org/D14-1181}
\BIBentrySTDinterwordspacing

\bibitem{hou2021causal}
S.~Hou, W.~Xu, J.~Chai, C.~Wang, W.~Zhuang, Y.~Chen, H.~Bao, and Y.~Wang, ``A
  causal convolutional neural network for motion modeling and synthesis,''
  2021.

\bibitem{shi2020motionet}
M.~Shi, K.~Aberman, A.~Aristidou, T.~Komura, D.~Lischinski, D.~Cohen-Or, and
  B.~Chen, ``Motionet: 3d human motion reconstruction from monocular video with
  skeleton consistency,'' \emph{ACM Transactions on Graphics (TOG)}, vol.~40,
  no.~1, pp. 1--15, 2020.

\end{thebibliography}

\end{document}